\documentclass[epj]{svjour}
\usepackage{amsmath}
\usepackage{amsmath}
\usepackage{amssymb}
\usepackage{amsfonts}
\usepackage{graphicx}
\usepackage{float}
\usepackage{wrapfig}
\usepackage{subfigure}
\usepackage{bm}
\usepackage{units}
\usepackage{grffile}
\usepackage{color}
\usepackage{hyperref}

\def\MB#1{{\textcolor{black}{ #1}}}    % Michele

\newcommand{\be}{\begin{equation}}
\newcommand{\ee}{\end{equation}}

\newcommand{\bu}{\boldsymbol{u}}

\newcommand{\bhu}{\boldsymbol{\hat{u}}}

\newcommand{\bF}{\boldsymbol{f}}

\newcommand{\bk}{\boldsymbol{k}}

\newcommand{\bx}{\boldsymbol{x}}

\newcommand{\Lcal}{\mathcal{L}}
\newcommand{\Ocal}{\mathcal{O}}

\begin{document}

\title{Classifying Turbulent Environments via Machine Learning}

%%%%%%%%%%%%%%
\author{Michele Buzzicotti\inst{1} \and Fabio Bonaccorso\inst{1}}
% \offprints{}          % Insert a name or remove this line
\institute{Department of Physics \& INFN, University of Rome `Tor
Vergata', Via della Ricerca Scientifica 1, 00133 Rome, Italy}
\abstract{The problem of classifying turbulent environments from partial observation is key for some theoretical and applied fields, from engineering to earth observation and astrophysics, e.g.  to precondition searching of optimal control policies in different turbulent backgrounds, to predict the probability of rare events and/or to infer physical parameters labelling  different turbulent set-ups.  To achieve such goal one can use different tools depending on the system's knowledge and on the quality and quantity of the accessible data. In this context, we assume to work in a model-free setup completely blind to all dynamical laws, but with a large quantity of (good quality) data for training. As a prototype of complex flows with different attractors, and different multi-scale statistical properties we selected 10 turbulent `ensembles' by changing the rotation frequency of the frame of reference of the 3d domain and we suppose to have access to a set of partial observations limited to the instantaneous kinetic energy distribution in a 2d plane, as it is often the case in geophysics and astrophysics. We compare results obtained by a Machine Learning (ML) approach consisting of a state-of-the-art Deep Convolutional Neural Network (DCNN)  against Bayesian inference which exploits the information on velocity and enstrophy moments. First, we discuss the supremacy of the ML approach, presenting also results at changing the number of training data and of the hyper-parameters. Second, we present an ablation study on the input data aimed to perform a ranking on the importance of the flow features used by the DCNN, helping to identify the main physical contents used by the classifier. Finally, we discuss the main limitations of such data-driven methods and potential interesting applications.}

\maketitle
\section{Introduction}
\label{sect:Introduction}
%Fig 1
\begin{figure*}[h]
    \centering
    \includegraphics[scale=.22]{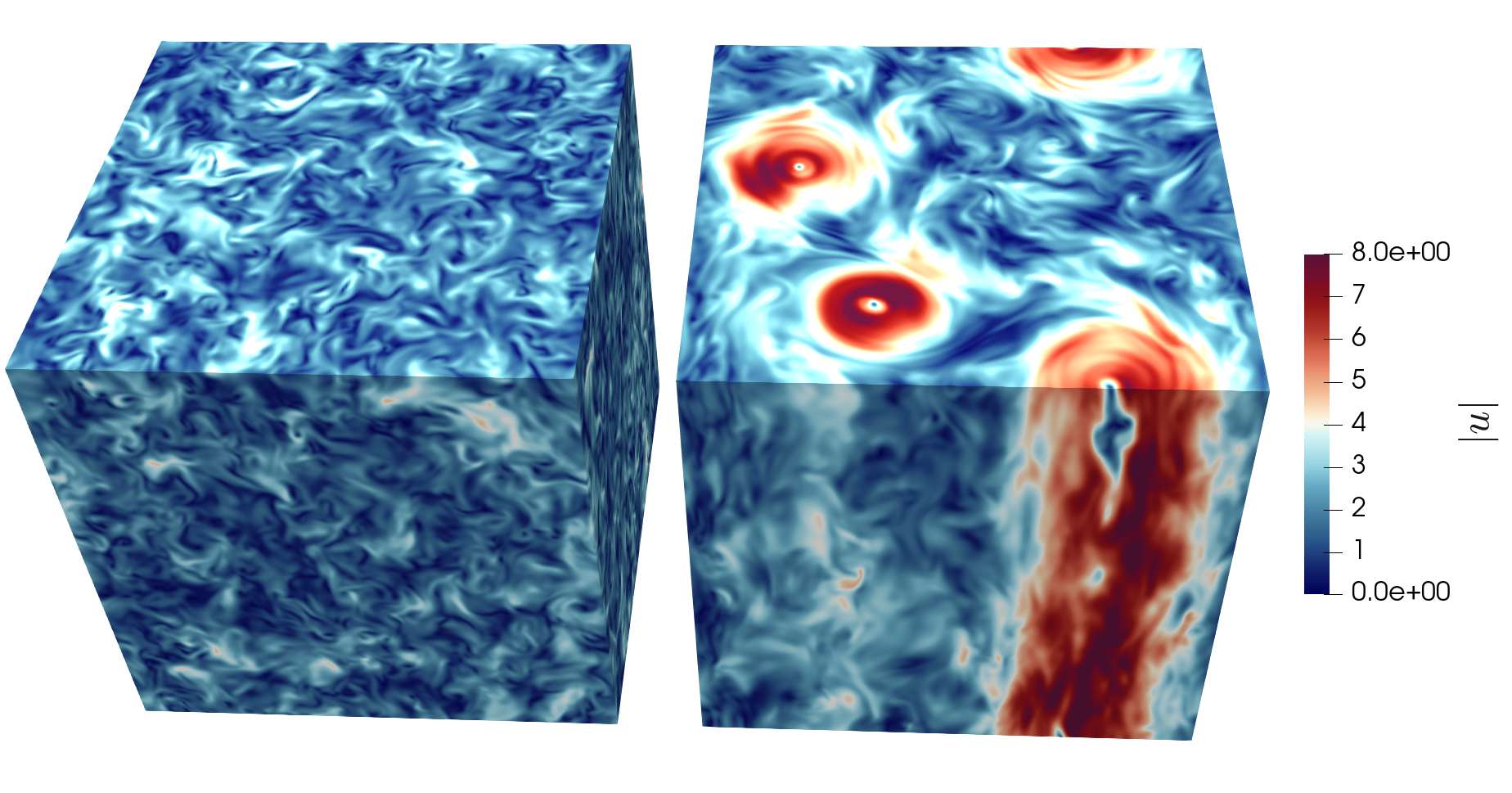}
    \caption{3D rendering of a turbulent flow under rotation at two different values of rotation rate, namely $\Omega=4$ on the left side and $\Omega=10$ in the right panel. In the first case the rotation frequency is small enough that the flow develops a dynamics similar to the standard 3D non-rotating case, with forward cascade of energy and small scales fluctuations. In the second case instead rotation is large enough to produce an inverse energy transfer that give rise to large-scales structures almost constant in the rotation direction.}
    \label{fig:figure1}       % Give a unique label
\end{figure*}

\MB{Extracting information about the statistical context from partial  measurements of a turbulent  system is a fundamental problem in fluid dynamics. Unless the data available are large and of very high quality, any linear regression trying to statistically distinguish different turbulent set up from limited data is destined to fail. This is evident if you need, e.g.,  to estimate the turbulent intensity, given by a combination of  the velocity root mean square, viscosity and flow correlation length, out of sparse and temporal measurements, and without any prior on the dissipative properties \cite{corbetta2021deep}. Similarly, many turbulent set-ups are subjected to transitions at changing control parameters or boundary conditions, developing macroscopically similar behaviour but different subtle multi-scale statistical properties and/or transient phases \cite{alexakis2018cascades,frisch1995turbulence,Pope00,davidson2011voyage,duraisamy2019turbulence}. As a result, classifying  turbulent scenarios from spot measurements is a key problem to reduce the complexity of predicting the flow evolution, to select optimal policies, pre-trained in different set-ups, for optimal navigation \cite{biferale2019zermelo,novati2019controlled,orzan2022optimizing,garnier2021review,colabrese2017flow,reddy2016learning,reddy2018glider} and/or estimating the probability of extreme events \cite{frisch1995turbulence,scatamacchia2012extreme,buzzicotti2021inertial,biferale2016coherent,buaria2020self,yeung2015extreme}.
Similarly in the context of data-driven turbulence modelling of non-stationary and transient flows, an instantaneous classification of the environment would be beneficial to perform an ad-hoc fine-tuning of the closure model depending on the specific local turbulence condition, e.g. high/low shear, stable/unstable conditions or  strong/weak updrafts, such as to cure the `curse-of-dimensionality' and improve the training performances \cite{maulik2019subgrid,pathak2020using,kochkov2021machine,biferale2019self}. As a prototype of complex flows with different dynamical attractors, and different multi-scale
statistical properties we selected 10 turbulent `ensembles' by changing the rotation frequency of the
frame of reference of a 3d domain and we supposed to have access to a set of partial observations limited to the instantaneous kinetic energy distribution in a 2d plane, as it is often the case in geophysics and astrophysics~\cite{vallis2017atmospheric,pedlosky1987geophysical,akyildiz2002wireless,kalnay2003atmospheric,buzzicotti_ocean_2021,carrassi2008data,brunton2016discovering,bocquet2019data}. Increasing rotation rate, turbulence undergoes global and local changes, concerning  the development of quasi-2d columnar structures, extreme non-Gaussian small-scale anisotropic fluctuations, long-range wave-wave correlations \cite{smith1999transfer,mininni2009helicity,biferale2021rotating,di2020phase}, as a result, it can be considered a paradigmatic test-bed where spatial and Fourier features have different fingerprints in the statistical distributions.}  \\
\noindent
\MB{Classification methods can be split in two different classes. The equation-driven tools, where the data analysis is enhanced by the physical knowledge of the underlying dynamical laws \cite{zou1992optimal,ruiz2013estimating,di2018inferring}, and the purely data-driven methods, based on the statistical analysis of the available data. Some attempts to classify turbulence using equation-driven tools have been developed in the recent years, using Fourier spatio-temporal decomposition \cite{di2015spatio} or Nudging \cite{di2018inferring,di2020synchronization,buzzicotti2020synchronizing}. However, all these techniques on top of the knowledge of the equation-of-motion require access to important numerical resources to run digital twins of the observed system.
In this work we discuss the  model-free approaches, where we are completely agnostic of the system dynamical laws and we aim to implement purely data-driven tools trained on a large quantity of good quality data. }
%Fig 2
\begin{figure*}[h]
\centering
\includegraphics[scale=.5]{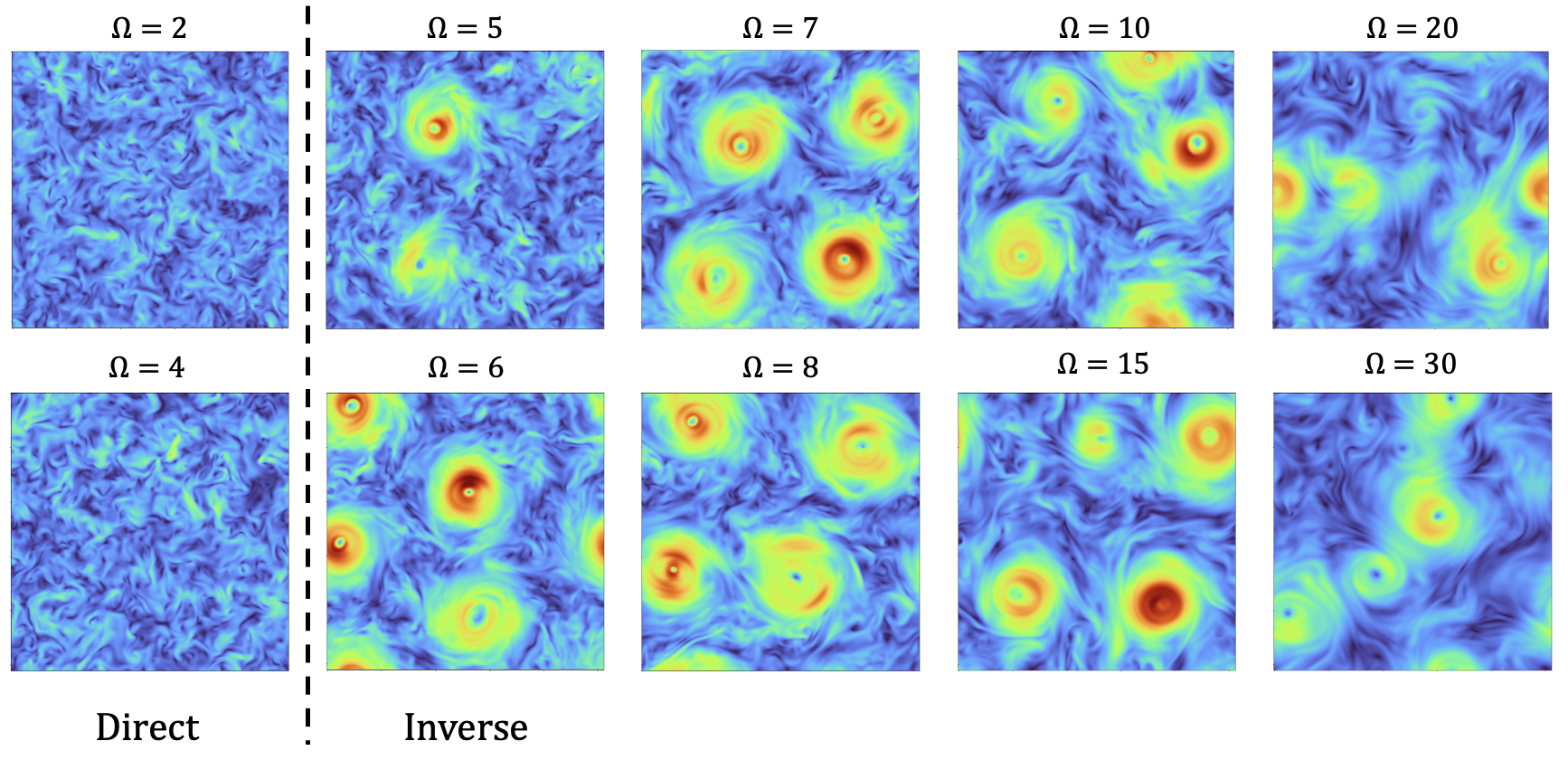}
\includegraphics[scale=.5]{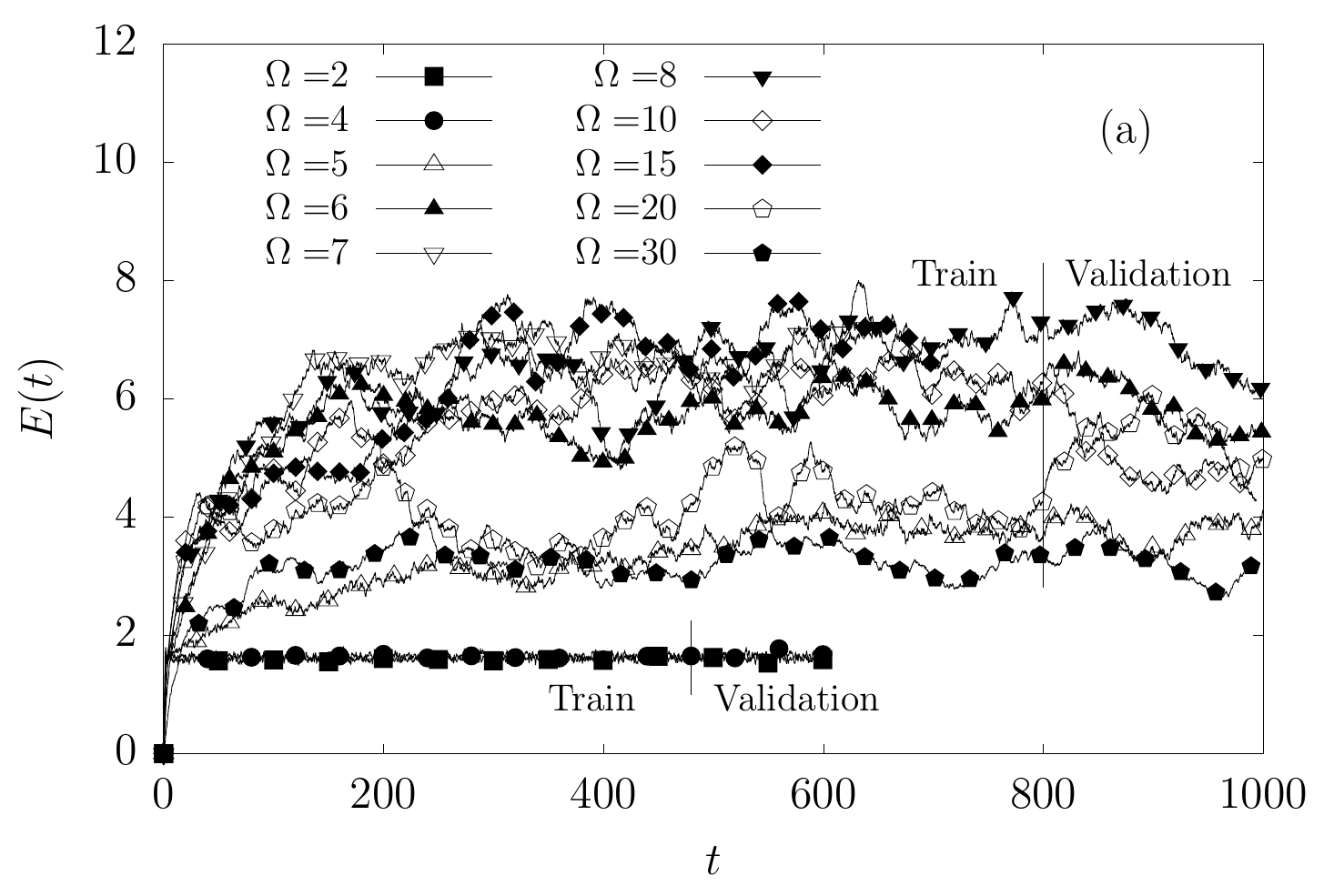}
\includegraphics[scale=.5]{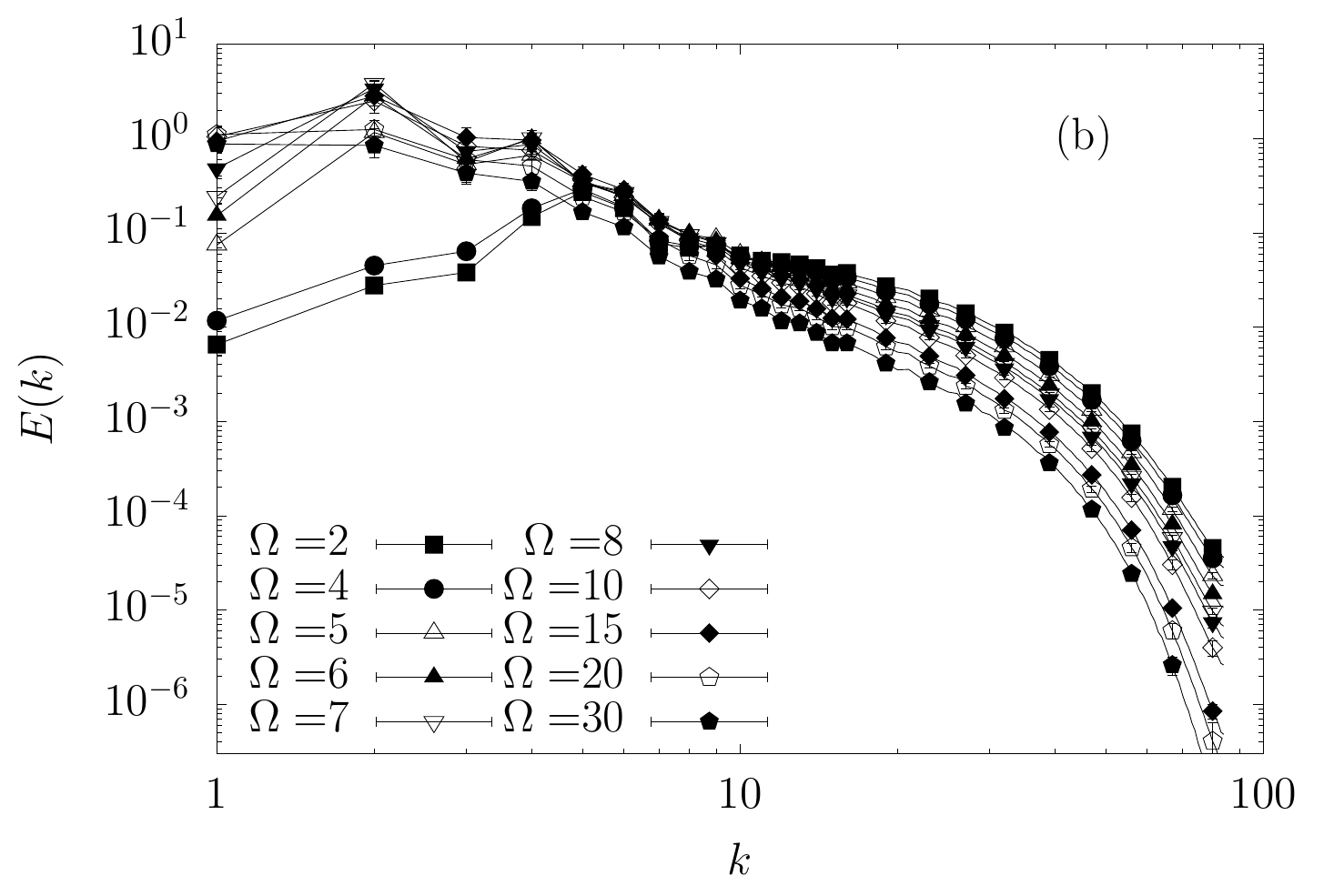}
\caption{Visualization of velocity magnitude on 2d planes extracted from a 3d domain, in the stationary regime for the 10 different turbulent `ensembles' used in the regression problem. The fields at $\Omega \le 4 $ are in the direct cascade regime, while increasing the rotation rate we observe the transition to the inverse energy cascade regime. (Bottom row panel `a') Energy evolution for the 10 different simulations for different $\Omega$, the dashed lines split the data used for the training and for validation sets. (Bottom row panel `b') Energy spectra as a function of the wavenumbers for the same set of simulations. The spectra are averaged on time in the stationary regime, errorbars indicate the standard deviation and are generally inside the symbols size.}
\label{fig:figure2}
\end{figure*}
\MB{\noindent In this paper, we attack this general {\it inverse problem}  following recent progresses in Machine Learning (ML) augmented fluid dynamics, \cite{brenner2019perspective,DuraisamyARFM2019,vinuesa2022enhancing,goodfellow2016deep,brunton2019data,brunton2020machine,biferale2019zermelo,buzzicotti2021reconstruction,brajard2019combining,borra2021using,corbetta2021deep}, by training a state-of-the-art  Deep Convolutional Neural Network (DCNN) in a supervised way, \cite{krizhevsky2012imagenet,he2016deep,ronneberger2015u,redmon2016you}, thanks to a large dataset build up with the use of high performance Direct Numerical Simulations (DNS) and accessible at the webpage \url{http://smart-turb.roma2.infn.it}. We compare the non-linear ML regression against a Bayesian baseline, based on the estimation of velocity and vorticity flow moments, in order to have comparison with both large- and small-scale physics inputs. Furthermore, we critically analyze the performance of the DCNN at changing the amount of training data and the number of optimized hyper-parameters. Finally, we try to {\it open-the-black-box} by performing an ablation study on the input data aimed to rank the flow features used by the DCNN, and to help to identify the main
physical contents used by the classifier. Such `inverse-engineering' approach, identifying key degrees-of-freedom for the classification can also have important applications concerning active control to favour or contrast flow transitions in other general set-ups.   
The paper is organized as follow. In Sect.~\ref{sect:background} we give details on the physics of turbulent flows and on the regression problem we are focusing on. In Sect.~\ref{sect:Dataset} we describe the datasets used. In Sect.~\ref{sect:Methods} we describe the methodologies implemented, the DCNN and the BI. Results are presented in Sect.~\ref{sect:Results}. \MB{In Sect.~\ref{sect:Occlusion} we present the ablation study on the input data,} and in Sect.~\ref{sect:Conclusions} we draw our concluding remarks.}
%Fig 3
\begin{figure*}[h]
\centering
\includegraphics[scale=.6]{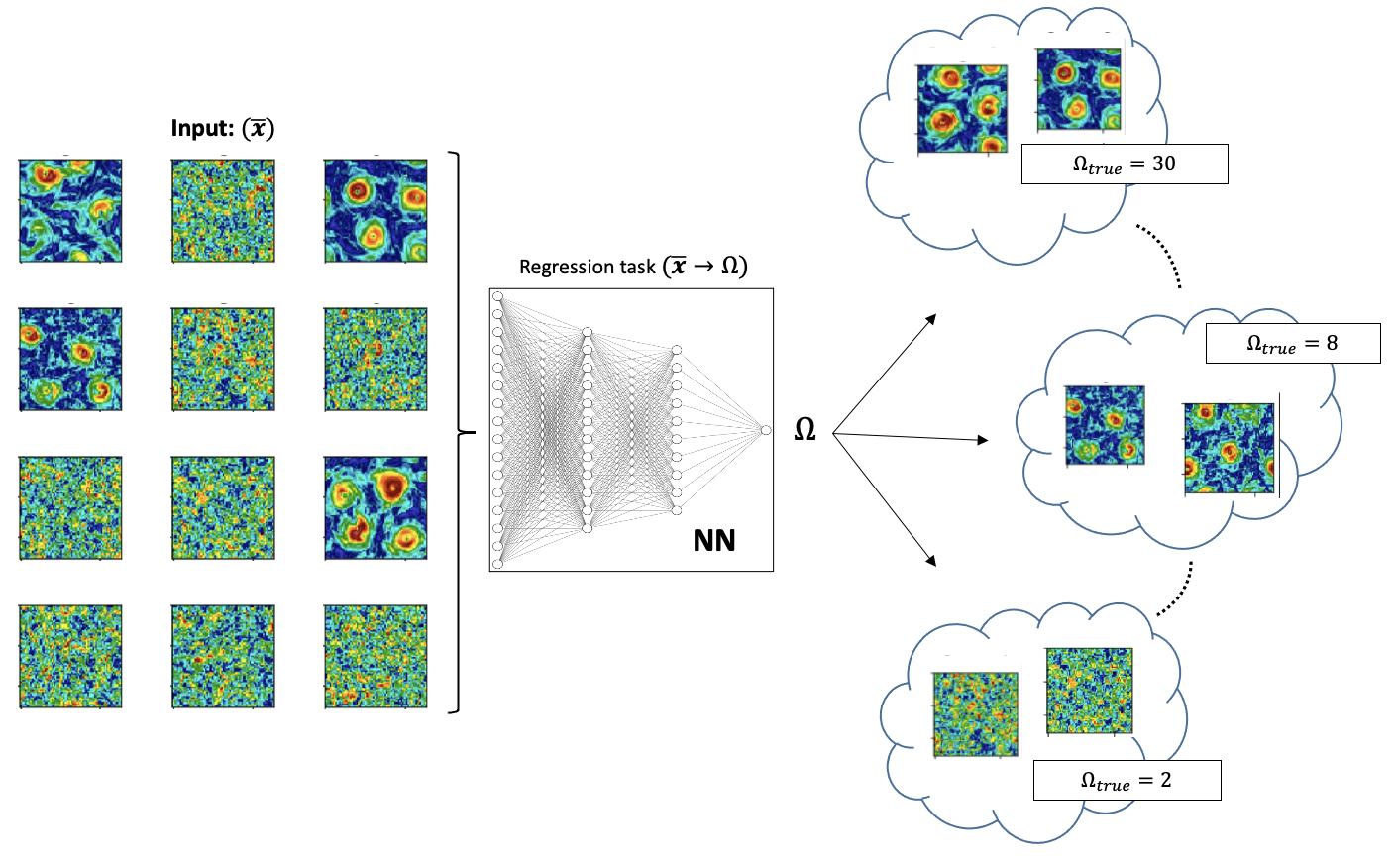}
\centering
\caption{Visual representation of the DCNN input-output setup used in the regression problem. The neural network is trained to map a plane of velocity module extracted from 3d data to the rotation frequency of the reference frame.}
\label{fig:figure3}       % Give a unique label
\end{figure*}

\section{Physics Background} 
\label{sect:background}
Before moving towards the details of the regression problem let us discuss the physical background of turbulence under rotations.
\noindent
Rotating turbulence is described by the Navier-Stokes equations (NSE) equipped with the Coriolis force,
\begin{equation}
\label{eq:nse}
\begin{cases}
\partial_t \bu + \bu \cdot \nabla \bu +2 \Omega\hat{z} \times {\bu}
= - \nabla p + \nu\Delta\bu+ \bF \\
\nabla \cdot \bu = 0,
\end{cases}
\end{equation}
here $\nu$ is the kinematic viscosity, $2{\Omega\hat{z}} \times {\bu}$ is the Coriolis force produced by rotation, and $\Omega$ is the frequency of the angular velocity around the rotation axis $\hat{z}$. The fluid density is constant and absorbed into the definition of pressure $p$. $\bF$, represents an external forcing mechanism, see Sect.~\ref{sect:Dataset} for details on the type of forcing used here. The relative importance of non-linear (advection) with respect to linear terms (viscous dissipation) is given by the Reynolds number, $Re =\varepsilon_f^{1/3} k_f^{-4/3}/\nu$, where $\varepsilon_f$ is the rate of energy input at the forcing scale $L_f \sim 1/k_f$. 
Rotation introduces a second non-dimensional control parameter, the Rossby number, which represents the ratio between rotation frequency and the flow characteristic time scale at the forcing wavenumber:
\be
Ro = \frac{(\varepsilon_f k_f^2)^{1/3}}{\Omega}.
\ee
While it is well known that increasing $Re$ leads to transitions to different turbulent regime, the effects of varying $Ro$ are more subtle and the results of contrasting flow features, including wave-wave interactions, large-scale energy condensation in quasi-2d structures, small-scale anisotropic intermittency \cite{biferale2016coherent}. In the limit of large $Ro$ or small $\Omega$, standard 3d homogeneous and isotropic turbulent dynamics is observed, on the other hand, increasing $\Omega$, $Ro\le 1$, the flow becomes  almost bidimensional \cite{deusebio2014dimensional,marino2013inverse,marino2014large,buzzicotti2018energy,buzzicotti2018inverse,di2020phase}, moving with uniform velocity along the direction parallel to the rotation axis, i.e. $\hat{z}$. Simultaneously, at high rotation rates, the flow develops coherent/large-scales vortex like structures that live on the horizontal plane, $(x,y)$, as can be seen in the  renderings of Fig.~\ref{fig:figure1}. These structures are fed by an inverse energy cascade that transports energy from the injection scales  towards the largest scales available in the physical domain \cite{smith1999transfer,seshasayanan2018condensates,van2020critical}. \MB{In Fig.~\ref{fig:figure2} we present a visualization of our problem: classifying the turbulent environments from one realization of the velocity module on 2d planes extracted from 3d simulations at varying the reference rotation rate $\Omega$. From the visual comparison of the 2d planes for different $\Omega$, (from $\Omega=2$ to $\Omega=30$), and the results shown for the spectra and for the temporal evolution of the total kinetic energy (panels `a' and `b') one can immediately understand how it is not so difficult to classify the systems in two rough classes, strong/weak rotations. On the other hand,  it looks much harder to distinguish different scenarios inside one of the two rough classes. The aim of the paper is to find a tool capable to disentangle also the subtle correlations that distinguish among all turbulent ensembles.
We aim to solve the inference problem with the new data analyses paradigms proposed in computer vision such as a non-linear classifier based on DCNN~\cite{janai2020computer,maron2020learning,pidhorskyi2018generative,tan2020efficientdet,chouhan2020applications,cheng2021fashion,feng2019computer,pathak2016context}.} 
We imagine  to have only a partial knowledge of the system, and in particular, for the sake of a possible realistic  implementation,  we make the four following assumptions:
\begin{itemize}
    \item Time evolution is not accessible 
    \item We have access only to one scalar measurement, e.g. the velocity amplitude, $|\bu|$      
    \item We can observe the system on a 2d plane and not on the full 3d domain
\end{itemize}
%The first limitation is the most stringent, and indeed  we don't know of any previous attempt to infer rotation rate without using temporal information.  %Indeed, up to our knowledge, there are not previous attempts based on Machine Learning techniques to attack the problem without any knowledge on the equations of motion.
\MB{As a baseline, we have repeated the same investigation using a Bayesian Inference (BI) analysis, in order to compare the `supremacy' of the black-box DCNN with a tool based on correlations with  physically meaningful observables, i.e.  moments of the velocity field and of its derivatives.} 
Let us notice that we have also attempted to perform the same classification, using the unsupervised Gaussian Mixture Models (GMM) classification~\cite{rasmussen1999infinite,reynolds2009gaussian,HeGMM2011}. However we did not find any relation between the unsupervised classification and the rotation value, so it has not been included in the discussion with DCNN and BI.
%Fig 4
\begin{figure*}[h]
\centering
\includegraphics[scale=.5]{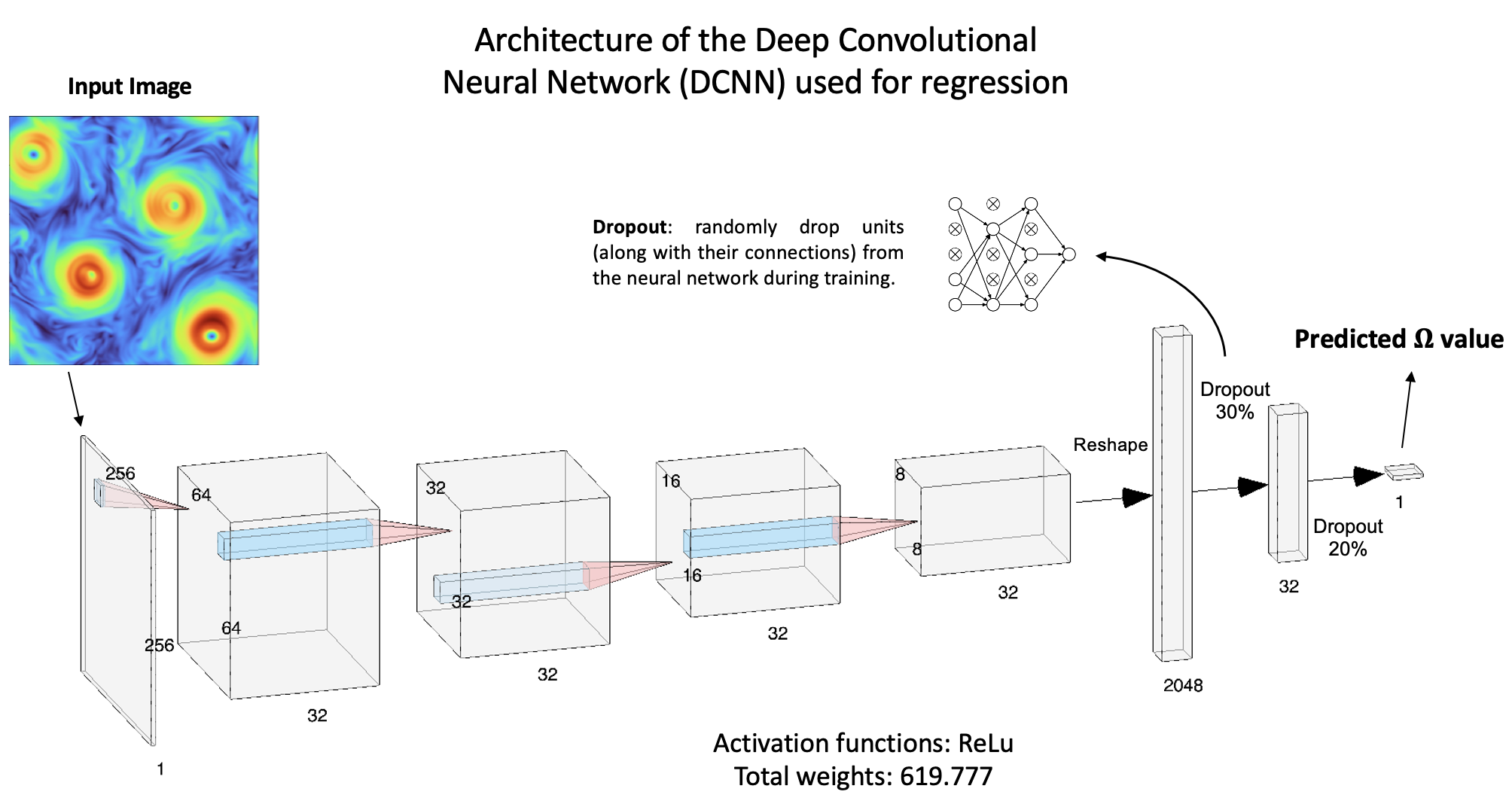}
\caption{Details of the Deep Convolutional Neural Network implemented in this work. \MB{The network starts with four convolutional layers that map the input plane to 32 different 8x8 images of aggregated features. The second part of the network is composed by two fully connected layers that give in output the rotation prediction, $\Omega_{pred}$. All layers are followed by a ReLu activation function, a dropout of $30\%$ and $20\%$ follows respectively the two last fully connected layers.}}
\label{fig:figure4}       % Give a unique label
\end{figure*}

\section{Dataset}
\label{sect:Dataset}

\subsection{Numerical Simulations}
To generate the database of turbulent flows on a rotating frame, we have performed a set of Direct Numerical Simulations (DNS) of the NSE for incompressible fluid in a triply periodic domain of size $L=2\pi$ with a resolution of $N= 256$ gird-points per spatial direction. We used a fully dealiased parallel pseudospectral code, with time integration implemented as a second-order Adams-Bashforth scheme and the viscous term implicitly integrated. A linear friction term, $0.1 \Delta^{-1}\bu$, is added at wave-numbers with module $ |\bk|\le 2$ to reach a stationary state. At the same time, viscous dissipation, $\nu \nabla^{2} \bu$, is replaced by a hyperviscous term, $\nu \nabla^{4} \bu$, with  $\nu = 1.6\times10^{-6}$, to increase the inertial range of scales dominated by the non-linear energy transfer. 
The forcing mechanism, $\bF$, is a Gaussian process delta-correlated in time, with support in wavenumber space centered at $k_f=4$. 
The total energy time evolution for the 10 different simulations varying $\Omega$ is presented in Fig.~\ref{fig:figure2}(a), while in panel (b) of the same figure we show the energy spectra, $E(k)=\frac{1}{2} \sum_{ k \le|\bk| < k+1} |\bhu(\bk)|^2$, averaged over time in the stationary regime for the same set of 10 simulations.
From the energy spectra we can observe that while for $\Omega=2$ and $4$ (in the direct cascade regime) there is a depletion of energy at wavenumbers smaller than $|\bk|< k_f$, for $\Omega$ values above the transition the split cascade regime, also the small wavenumbers are filled in with energy driven by the inverse cascade, namely from the forcing to the largest system's scales.

\subsection{Dataset extraction}
%Fig 5
\begin{figure*}[h]
\centering
\includegraphics[scale=.39]{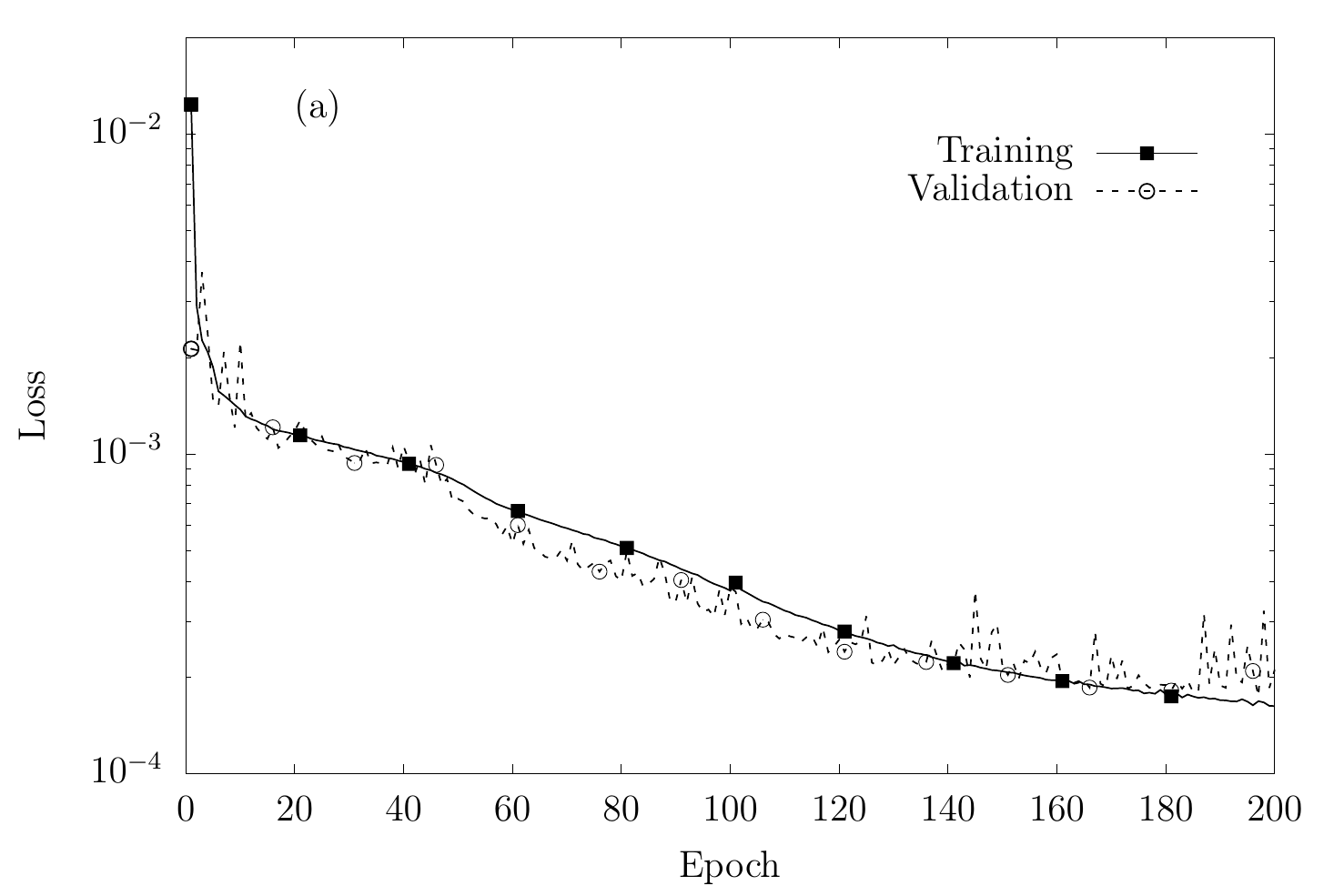}
\includegraphics[scale=.39]{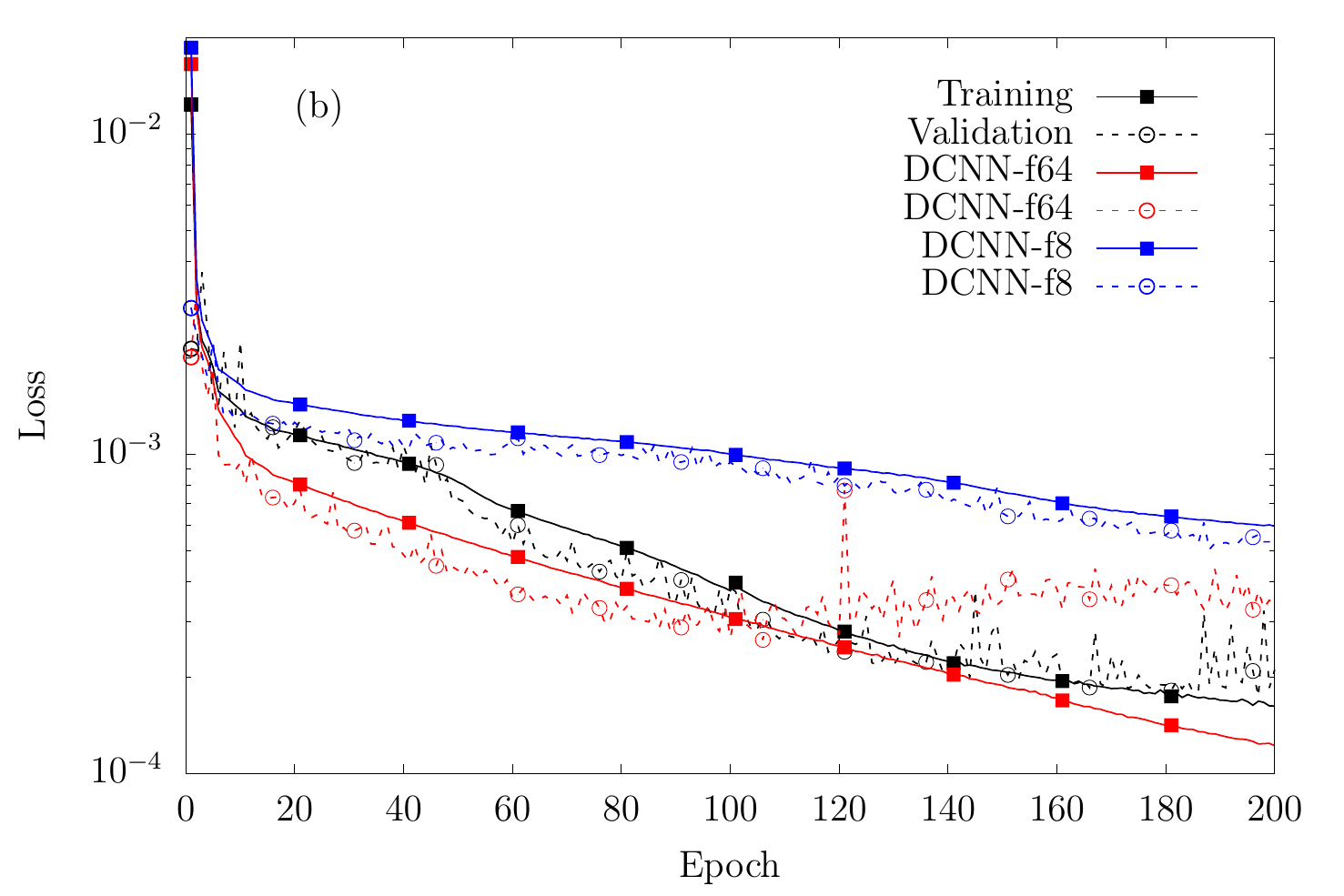}
\includegraphics[scale=.39]{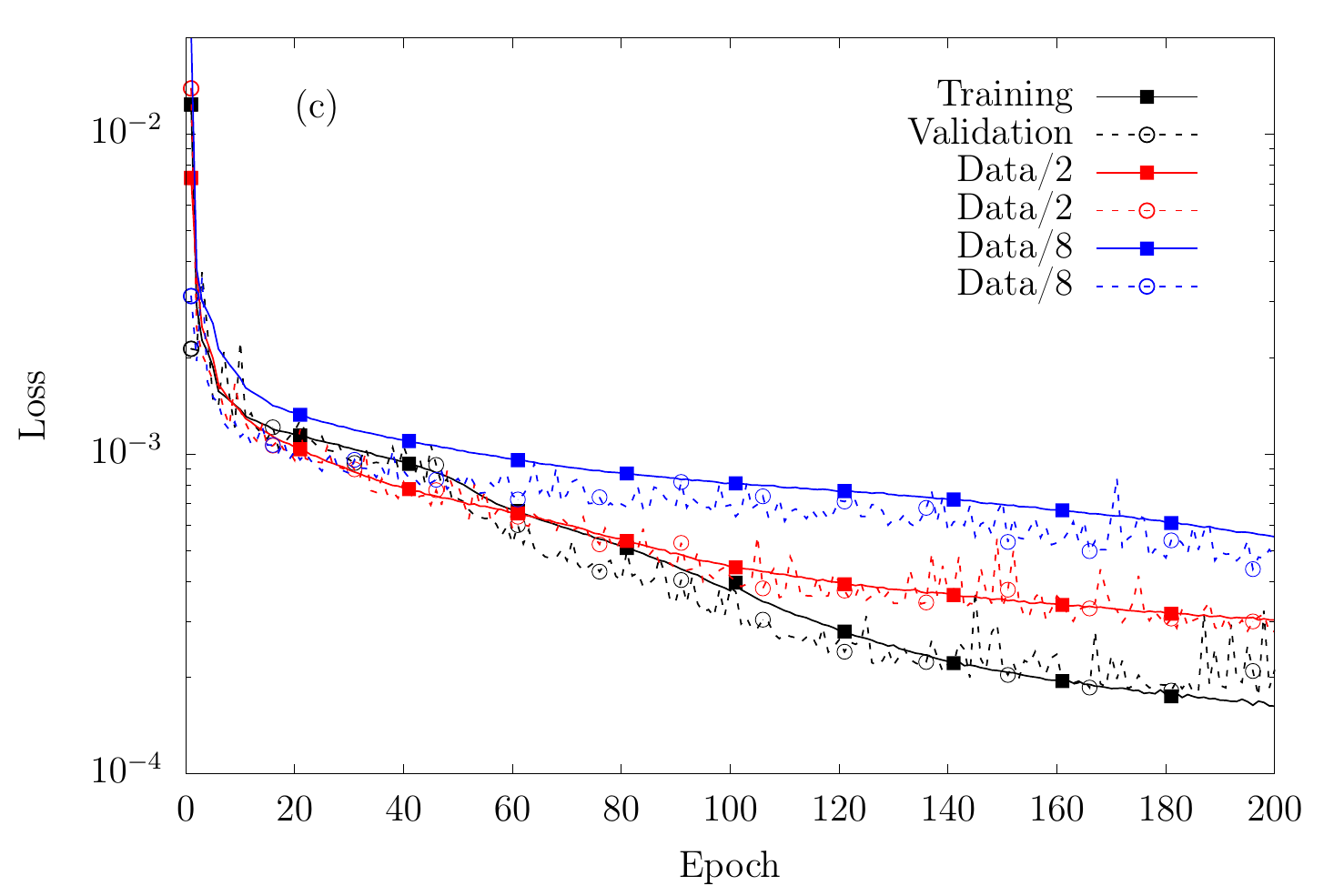}
\includegraphics[scale=.39]{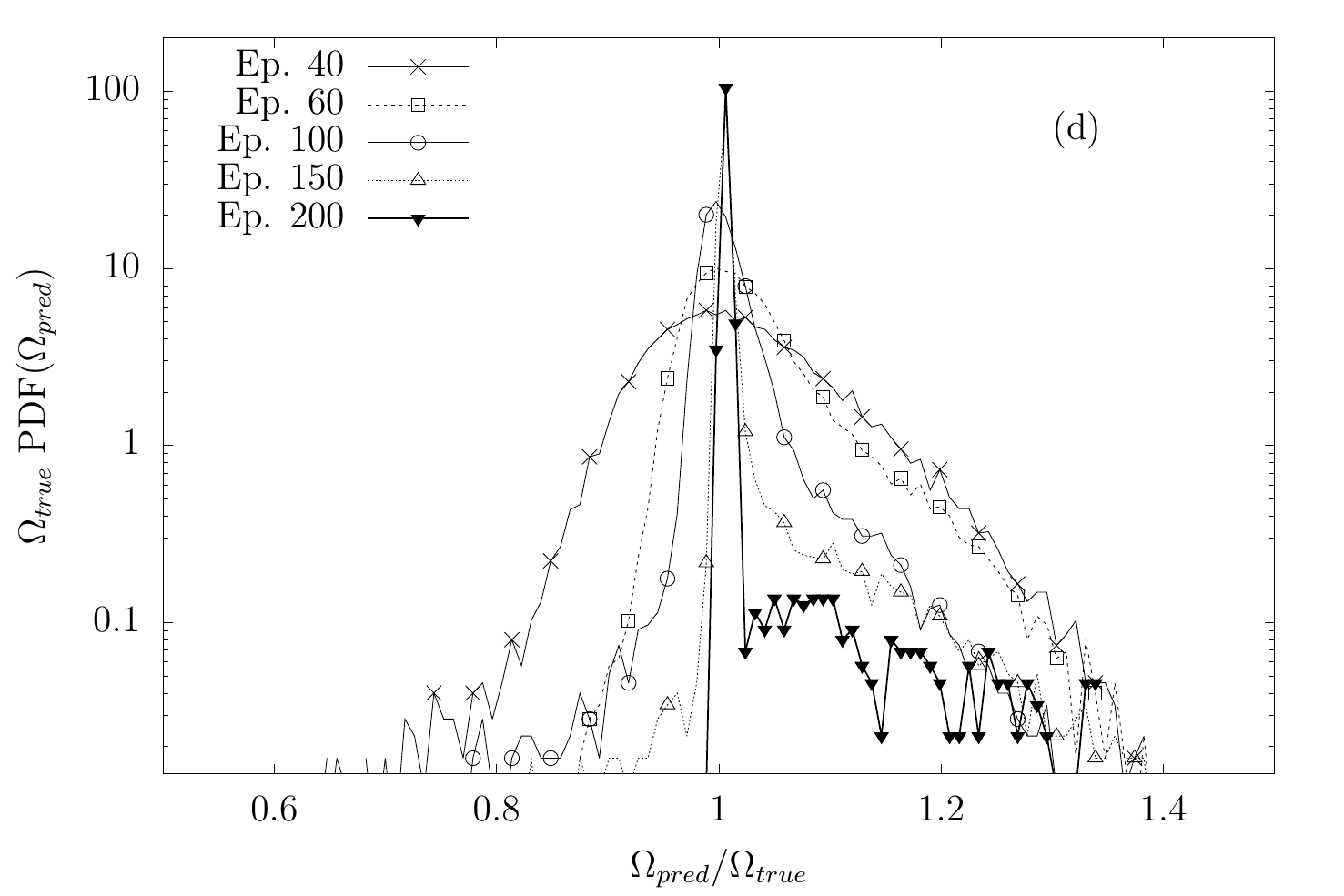}
\includegraphics[scale=.39]{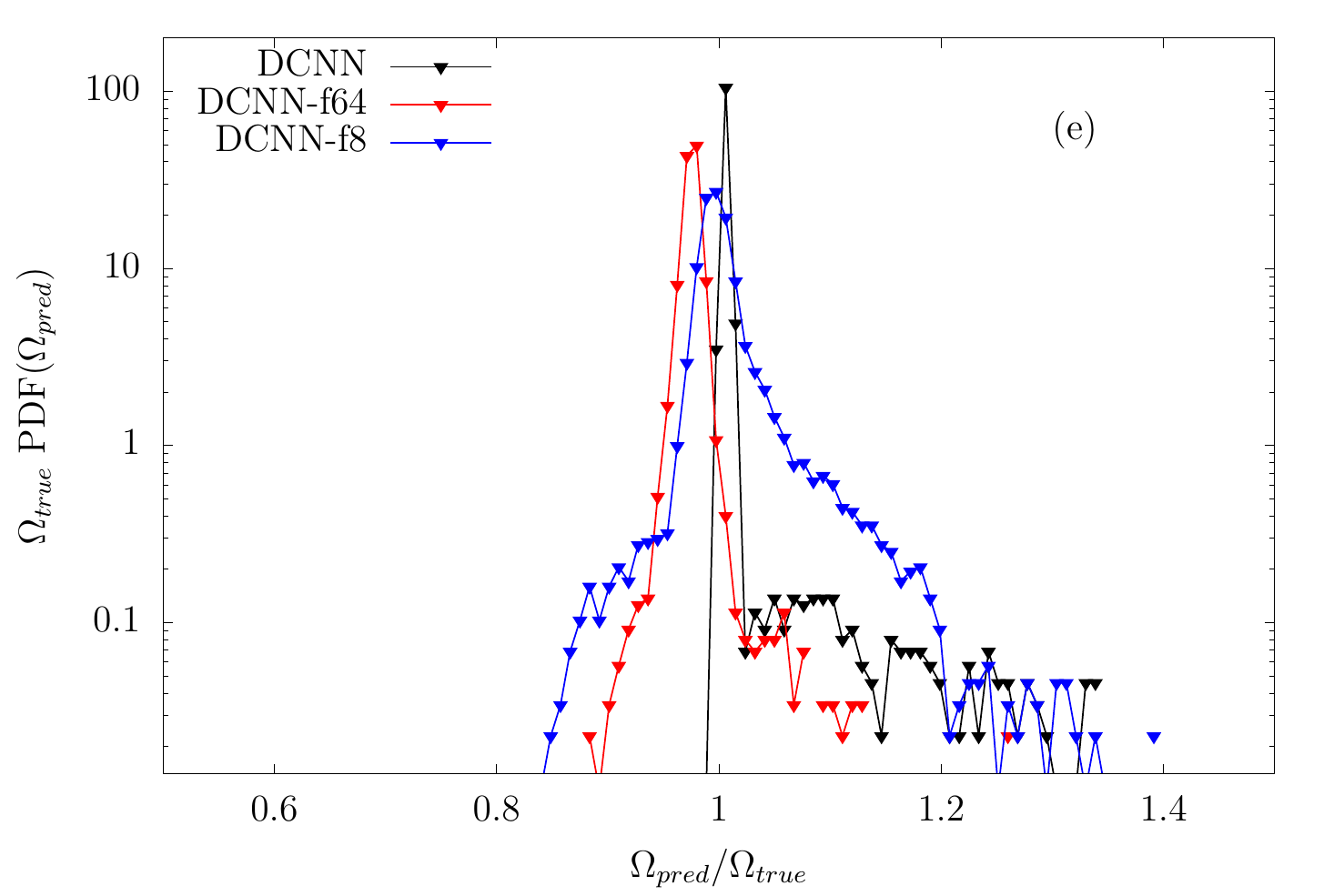}
\includegraphics[scale=.39]{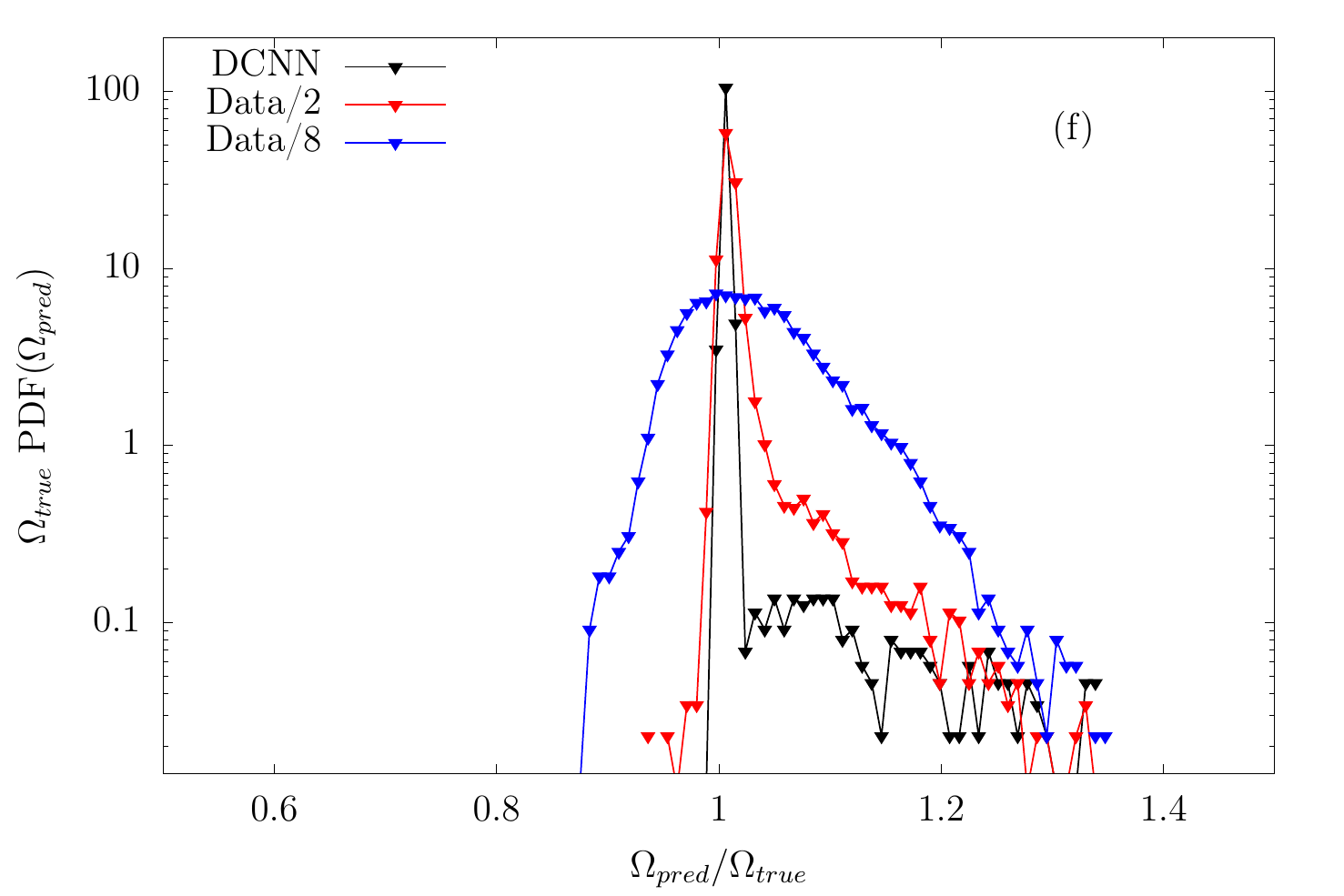}
\centering
\caption{(Panel a) Evolution of the Loss, $\mathcal{L}$, as a function of the epochs, measured on both the training set (solid line full squares) and the validation set (dashed line open circles), results are obtained with the DCNN introduced in Fig.~\ref{fig:figure4}. (Panel d) Probability Density Function (PDF), measured at different epochs, of the DCNN prediction normalized to the real value, for the case of $\Omega_{true}=20$. The PDFs are measured on the whole validation set. (Panels b and d) $\mathcal{L}$ and PDFs for the DCNN compared with the performances of two different neural networks with the same architecture of DCNN but with different weights. In particular, DCNN-f8 has 8 different filters in each convolutional layer and a total of $145201$  parameters, while DCNN-f64 has 64 filters in the convolutional layers and a total of $1403073$ trainable weights. (Panels c and f) Show the same results presented in panels (a) and (d) but reducing the size of the training set by a factor 2 (red lines) and by a factor 8 (blue lines) corresponding respectively to a total number of $20$k and $5$k training planes for each $\Omega$.}
\label{fig:figure5}       % Give a unique label
\end{figure*}

The ML training and validation datasets, accessible at the webpage \url{http://smart-turb.roma2.infn.it}, are extracted from the DNS described above, following~\cite{TurbRot}, such as to match all assumption discussed in the introduction (Sect.~\ref{sect:Introduction}), namely;
\begin{itemize}
    \item In order to vary the Rossby number, we performed 10 numerical simulations each with a different rotation frequency. Two simulations well below the transition in the direct cascade regime, $\Omega=2, 4$, three of them in the transition region, $\Omega=5,6,7$, and the remaining five well inside the split cascade regime with $\Omega=8,10,15,20,30$. 
    \item For each of the 10 simulations we have dumped a number of roughly $600$ snapshots of the full 3d velocity field with a temporal separation large enough to decrease correlations between two successive data-points (see Fig.~\ref{fig:figure2}(a) for the total energy evolution).
    %\item Each snapshot at the original resolution of $256^3$ grid points is downsized on a coarser grid of $64^3$ grid-points, after applying a Galerkin truncation in Fourier space where the maximum frequency is set to $|\bk| \le  32$.
    \item For each configuration, $64$ horizontal planes  $(x,y)$, (perpendicular to the rotation axis), are selected at different $z$ and for each of them we computed the module of the velocity field as; $|\bu|=\sqrt{u_x^2+u_y^2+u_z^2}$. 
    \item To increase the dataset, we shift each of the $64$ planes by choosing randomly a new center of the plane and using periodic boundary conditions, such as to obtain a total number of $50k$ planes for each $\Omega$ value.
\end{itemize}
The full dataset, in this way, is composed of half a million different velocity module planes, out of which the $80\%$ of them are used as training set while $20\%$ are used as validation set. To be sure that validation data were never seen during training, none of the planes used in the validation set is obtained from a random shift of a plane contained in the training set. Instead, the two sets are built using the same protocol but from different temporal realization of the flow splitting its time evolution in two separate parts as also displayed by the vertical dashed lines in panel (a) of Fig.~\ref{fig:figure2}.

%##########################################################

\section{Methods}
\label{sect:Methods}

\subsection{Deep Convolutional Neural Network (DCNN)}
Fig.~\ref{fig:figure3} schematizes the ML problem set-up. A plane of velocity module randomly extracted from the dataset of 10 simulations is given in input to a DCNN with the goal to infer the $\Omega$ value of the simulation from which the plane has been generated. The training is performed in a supervised setting, having pre-labelled all planes with their corresponding value, $\Omega_{true}$. As solution of the regression problem, the DCNN gives in output a real number corresponding to the predicted rotation value, $\Omega_{pred}$. In this way the inferring task is not limited to a predefined set of classes and can be generalized to estimate rotation values never observed during training. The network is trained to {\it minimize} the following loss function; 
\begin{equation}
  \Lcal =  \mathbb{E} \{ \left (\Omega_{true} - \Omega_{pred} \right )^2 \},
\label{eq:loss}
\end{equation}
which is the mean-squared-error between the target value $\Omega_{true}$ and the predicted value $\Omega_{pred}$. $\mathbb{E}$ is the mean over a mini-batch of 256 different planes.
To implement the DCNN we used TensorFlow \cite{abadi2016tensorflow} libraries with a Python interface, optimized to run on GPUs. \MB{Because of total dataset size, 245GB for the 500k different planes at a resolution of $256^2$ grid points, the training had to be parallelized over eight A100 Nvidia GPUs.}
\MB{As described in Fig.~\ref{fig:figure4}, the overall architecture is composed of four convolutional layers that encode the input plane of velocity module on $32$ different planes of size $8\times8$. This first part of the network is asked to aggregate the relevant features for the regression problem. After, the results for the aggregated features are reshaped into a single vector of $2048$ elements. The second half of the network, is composed by two fully connected layers that transform the features vector into one real number representing the network prediction, $\Omega_{pred}$. To reduce overfitting the training of these two final fully connected layers is performed using a dropout respectively of $30\%$ and $20\%$~\cite{srivastava2014dropout}. All layers are followed by a non-linear ReLu activation functions.}
Using a mini-batch of $256$ planes, and a training dataset of $N_{tra}=400k$ planes, each epoch consists of $1563$ back-propagation steps. 
\MB{The backpropagation has been performed with an Adam optimizer equipped with a Nesterov momentum \cite{dozat2016incorporating} using a decreasing learning rate starting from $10^{-4}$ with a scheduler that consists in a linear decrease of the learning rate up to reach $10^{-5}$ in 20 epochs and then to remain constant}. To achieve a good training with the learning protocol just introduced, we have normalized the input data to be in the $[0;1]$ range, namely we rescaled the module intensity of each gird-point by normalizing it to the maximum value calculated over the whole training set. The same normalization constant has been used to rescale the validation set. In the same way the training labels $\Omega_{true}$ have been rescaled by their maximum value in the training.
In Fig.~\ref{fig:figure5} panel (a) we show the evolution of the loss, $\Lcal$, measured on the training and validation data as a function of the number of epochs for the best model obtained in this work. The loss evolution does not show evidence of overfitting and converges after roughly $160$ epochs where the validation loss starts to saturate. 
In the bottom row, panel (d) of Fig.~\ref{fig:figure5}, we show the Probability Density Function (PDF) of predicting the correct rotation value during the different training epochs measured over all planes of the validation set. Here $\Omega_{true}=20$ is used, but similar behaviours are observed for all the other $\Omega$. At epoch $200$ the PDF becomes close to a delta-like distribution, suggesting very good accuracy of the network, with errors only every few cases over one-thousand, and within $30\%$ the correct value.   In the central column, panels (b,e) of Fig.~\ref{fig:figure5}, we compare the same results obtained varying the size of the DCNN, precisely the number of its trainable weights keeping fixed the architecture. Here the label DCNN-f8 and DCNN-f64 represent two neural networks that have respectively $8$ and $64$ different filters in each convolutional layer, instead of the $32$ as in the case of DCNN. The two networks have a total number of parameters of roughly $145$k for the smaller one and $1.4$M for the larger network. We can see that when the network is too small the accuracy decreases on both the validation and the training sets while if the network is too large there is some overfitting error on the validation set. Also the PDFs is optimal for the intermediate network labeled as DCNN. In the last column of Fig.~\ref{fig:figure5}, panels (c,f), we compare the DCNN results trained on smaller datasets. Here we can appreciate the quick deterioration of the prediction when reducing the quantity of the data used in the training. 

\subsection{Bayesian Inference}

The second statistical inference approach considered is the Bayesian Inference (BI)~\cite{aster2018parameter,efron2013250,mackay1992bayesian}. The idea is to infer properties of the flow assuming the knowledge of some underlying probability distributions. More specifically, the Bayes' Theorem tells us that the posterior probability, $P(\Omega_i | \Ocal)$, of observing $\Omega_i$ given a measure of another generic observable, $\Ocal$, can be estimated as follow; 
\begin{equation}
P(\Omega_i | \Ocal) = \frac{P(\Ocal |\Omega_i) P(\Omega_i)}{ P(\Ocal)} = \frac{P(\Ocal |\Omega_i) P(\Omega_i)}{ \sum_i P(\Ocal |\Omega_i) P(\Omega_i)}.
\label{eq:bayes}
\end{equation}
$P(\Omega_i)$ is the prior probability, in our case for all $\Omega_i$ equal to $0.1$ because the dataset is composed by the same number of planes for each of the 10 $\Omega_i$. $P(\Ocal | \Omega_i)$, instead, is the likelihood to get the measure $\Ocal$ of a generic observable for a fixed $\Omega_i$. 
Once known the posterior probability in Eq.~\eqref{eq:bayes}, a first Bayes estimation can be obtained as follows;
\begin{equation}
\overline{\Omega}_{Bay} = \sum_{i=1}^{10} P(\Omega_i|\Ocal) \Omega_i,
\label{eq:Omega_bay}
\end{equation}
which correspond to the mean value on the posterior probability. A second possible estimation can be done as the most likely $\Omega_i$ given $\Ocal$, namely;
\begin{equation}
\Omega^*_{Bay} = \max_{\Omega_i} P(\Omega_i|\Ocal) = \max_{\Omega_i} P(\Ocal|\Omega_i) P(\Omega_i),
\label{eq:Omega_bay_most}
\end{equation}
where in the last step the denominator of \eqref{eq:bayes} is neglected because it does not affect the maximization on $\Omega_i$.
\MB{For our problem, the two more relevant physical observable  are the total energy  and the total squared velocity gradients, representing  large-scale (Gaussian) and small-scale (non-Gaussian) features, respectively. Adding the information of flow time evolution would be beneficial to have a measure of the waves propagation inside the flow, but as discussed this is in general a much more complicated measure and we are not assuming to  rely on it. For this reason, in the sequel we will proceed consider $\langle u^2 \rangle$ and $\langle (\partial_x u)^2 \rangle$, where  $u = |\bu|$ is the module of the horizontal velocity and $\langle \cdots \rangle$ stands the average over the plane.} For each of these two observable we have calculated the likelihood $P(\Ocal | \Omega_i)$ from the $40k$  planes for each $\Omega_i$ of the training dataset. Namely we have measured $P( \langle u^2 \rangle | \Omega_i)$, $P( \langle (\partial_x u)^2 \rangle | \Omega_i)$ and the joint-likelihood, $P( \langle u^2 \rangle ,\langle (\partial_x u)^2 \rangle | \Omega_i)$ on the training set. 
%Fig 6
\begin{figure}[h]
%\sidecaption[h]
\includegraphics[scale=.6]{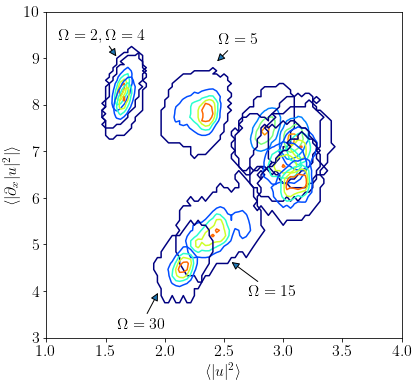}
\caption{Joint-likelihood, $P( \langle u^2 \rangle ,\langle (\partial_x u)^2 \rangle | \Omega_i)$, of the squared velocity module and its derivative along the $x$ direction, measured for the 10 different $\Omega$. The joint-PDFs are measured from the training set, and colours from dark-blue to red are proportional to the PDF value. The first cluster on the left consists of planes in the direct cascade, as indicate by the labels ($\Omega=2$ and $4$), the other clusters are composed by the overlapping of the remaining $\Omega$ values.}
\label{fig:figure6}       % Give a unique label
\end{figure}
\noindent
Measuring the same observable on each plane of the validation dataset, and knowing the likelihoods for each $\Omega_i$, we could estimate the posterior probability $P(\Omega_i | \Ocal)$ (Eq.~\eqref{eq:bayes}) and than infer $\Omega_{Bay}(\Ocal)$ (Eq.~\eqref{eq:Omega_bay}). 
Fig.~\ref{fig:figure6} shows the joint-likelihood of the mean energy and mean gradients. \MB{From this plot we can see that also combining large and small scales observables does not allow to have a clean statistical separation of the different $\Omega_i$ for all the 10 `ensembles' considered here.} 
In the next section we present and discuss results of $\Omega$ inference via DCNN and standard Bayesian approach.

%Fig 7
\begin{figure*}[h]
\centering
\includegraphics[scale=.5]{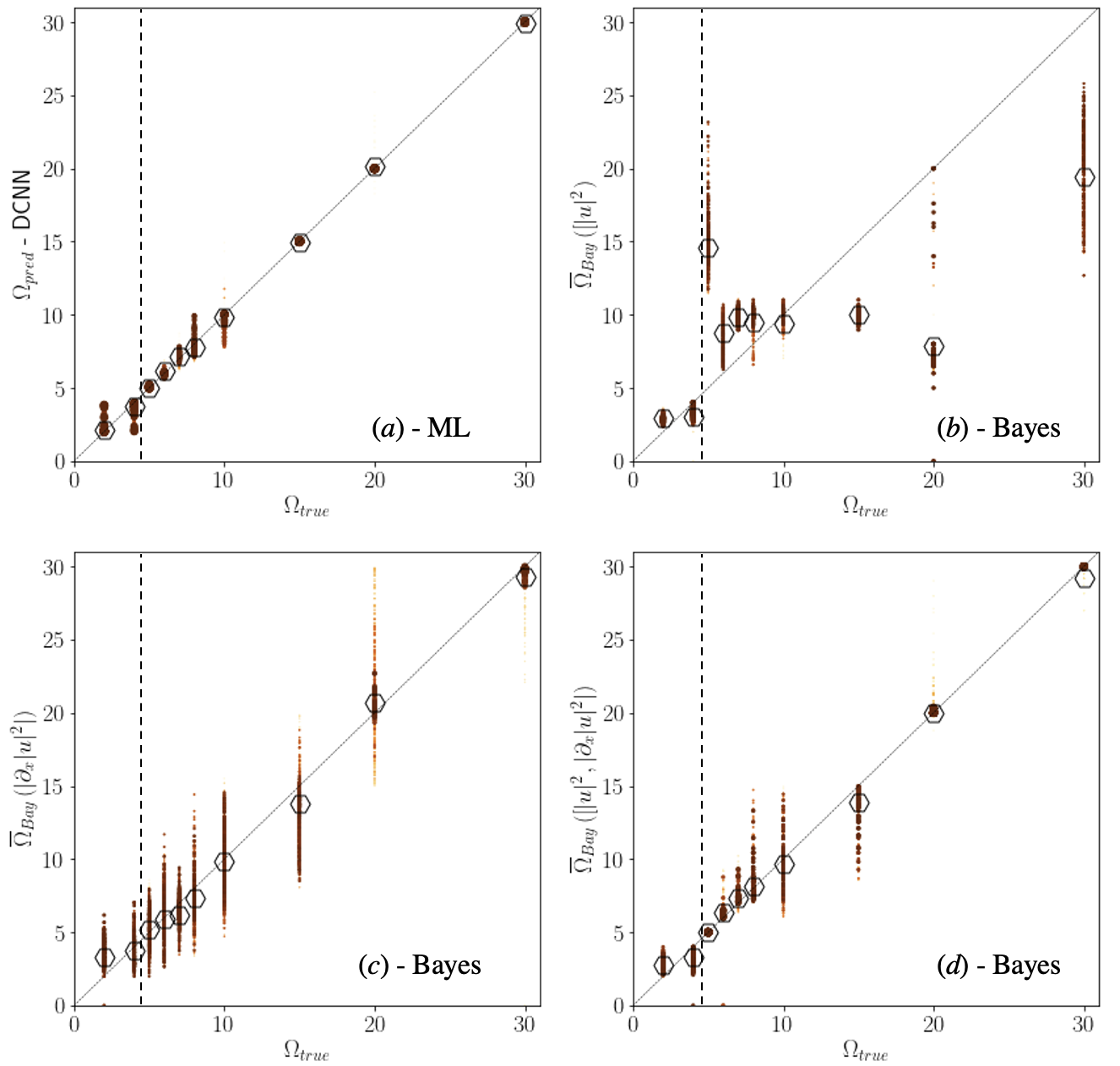}
\caption{Scatterplot of inferred vs real rotation value. The inference estimation is obtained with the ML approach, $\Omega_{pred}$ (panel a), or with BI, $\overline{\Omega}_{Bay}$, conditioning on the PDF of the velocity square module (panel b), of the velocity gradients (panel c) or with the joint PDF of velocity module and gradients (panel d). The diagonal lines indicate the identity function, while the red full circles dots are the predictions for each planes on the validation set, their size and color intensity is proportional to the density of points in a given area. The mean values over all predictions for all planes of each $\Omega_{true}$ are reported with the open hexagon symbols. The vertical dashed lines separate the two rough classes weak/strong rotation rates.}
\label{fig:figure7}       % Give a unique label
\end{figure*}

\section{Results}
\label{sect:Results}

%Fig 8
\begin{figure*}[h!]
\centering
\includegraphics[scale=.43]{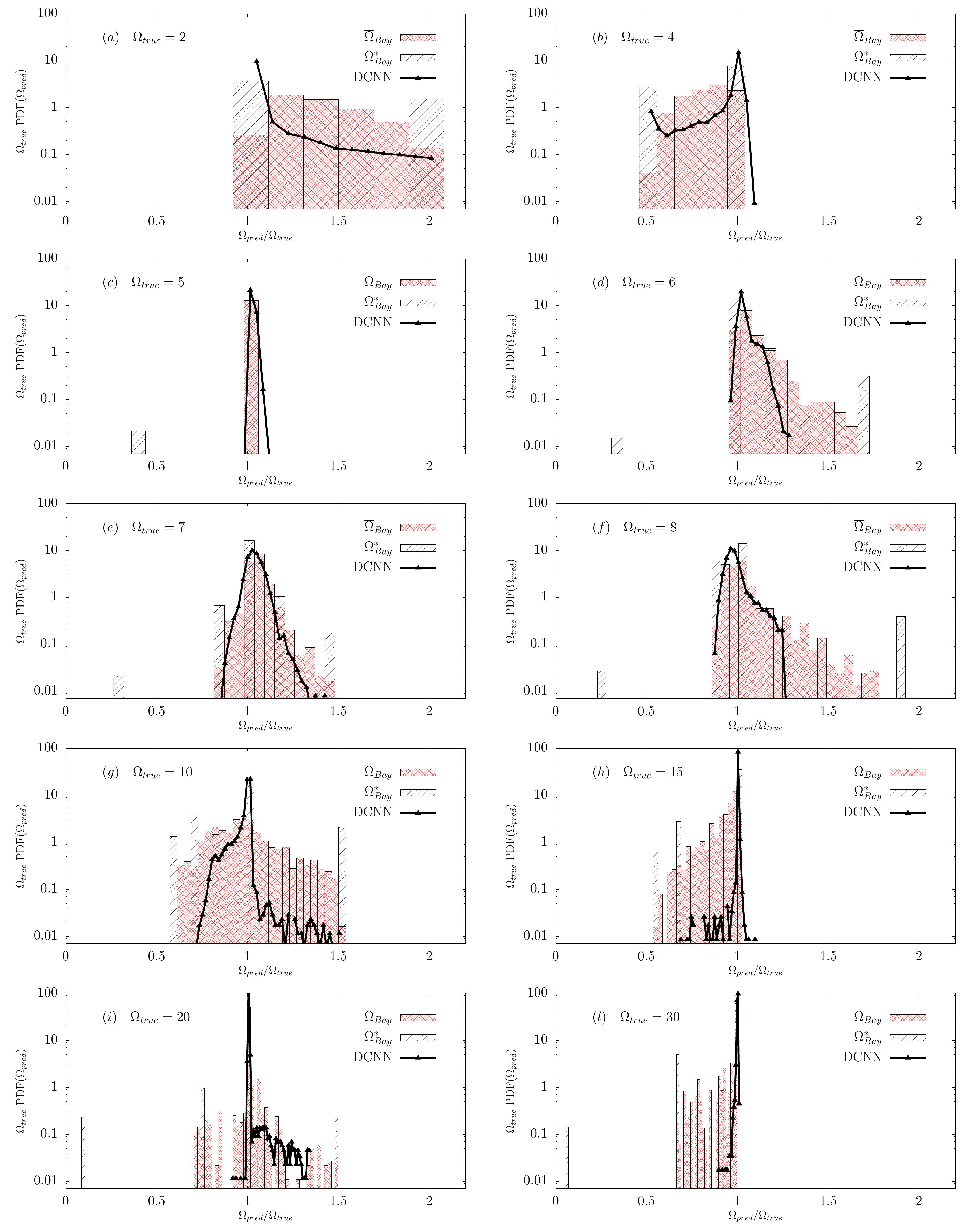}
\caption{PDF of rotation value predicted by the neural network, $\Omega_{pred}$ (DCNN, solid lines full triangles), compared with the mean, $\overline{\Omega}_{Bay}$ (red histograms), and most likely, $\Omega^*_{Bay}$ (black histograms),  Bayes inference using joint-likelihood $P( \langle u^2 \rangle ,\langle (\partial_x u)^2 \rangle | \Omega_i)$, on the validation set. For the sake of data presentation each of the PDF measured on the 10 different $\Omega$ values are reported in 10 different panels (from `a' to `l'). As indicated in the axis, the PDF are all normalized by their corresponding $\Omega_{true}$ indicate in the labels of each panel.}
\label{fig:figure8}       % Give a unique label
\end{figure*}
A first quantitative comparison of the two techniques is presented in Fig.~\ref{fig:figure7}, where in the four panels we report the scatterplot between the reference rotation value, $\Omega_{true}$ and the corresponding prediction obtained via the DCNN (panel a), and BI with the three different likelihoods, namely the one based on the mean energy (panel b), on the mean square gradients (panel c) and on the joint likelihood combining energy and gradients (panel d).
The four panels are all obtained with the same validation dataset of $10k$ planes for each of the 10 $\Omega_i$ values.
\MB{The first important observation  is that the DCNN prediction is surprisingly close to the correct values not only on average but also for each single sample. Indeed, from the top left panel of Fig.~\ref{fig:figure7} we can see that all points in the scatter plot fall well inside the hexagon symbol, just few exceptions are observed for the cases of $\Omega_{true}=8$ and $10$, and  for $\Omega_{true}=2$ and $4$. Let us stress that this result is far from being trivial and provides a first example where a ML data-driven technique succeed in the inferring of an hidden parameter from the analysis of real fully developed turbulent configuration. The reason of this success is double, the intrinsic potential of DCNN and the good quality of the dataset produced and implemented in the training of the network. Indeed as already shown a dataset reduction quickly leads to a deterioration of the training results, using a dataset up to eight times smaller we have seen that the DCNN predictions does not change on average but the points distribution around the correct values would be much larger, as shown for the case of $\Omega=20$ in the PDFs of Fig.~\ref{fig:figure5} panel (f).
The second  important consideration can be drawn by the observation of the BI results. Here, looking at panels (b-c-d) we can assess the ranking of different features for the classification task. When using only the energy, BI is able to distinguish among the two rough classes (below/above the energy transition) only, while it is completely blind to differences among the intermediate $\Omega$ values. In panel (c) we see that measuring small-scale features is more useful for the classification, at least on average. Panel (d) shows that combining small and large scales input improves the results, keeping the BI close to the ML approach.
Keeping adding different observable to condition the BI approach is clearly doomed of fail, because of the difficulties to have a large statistical sample to probe the conditional probability on a larger and larger dimensional space.}
%Figure 9
\begin{figure*}[h]
\centering
\includegraphics[scale=0.55]{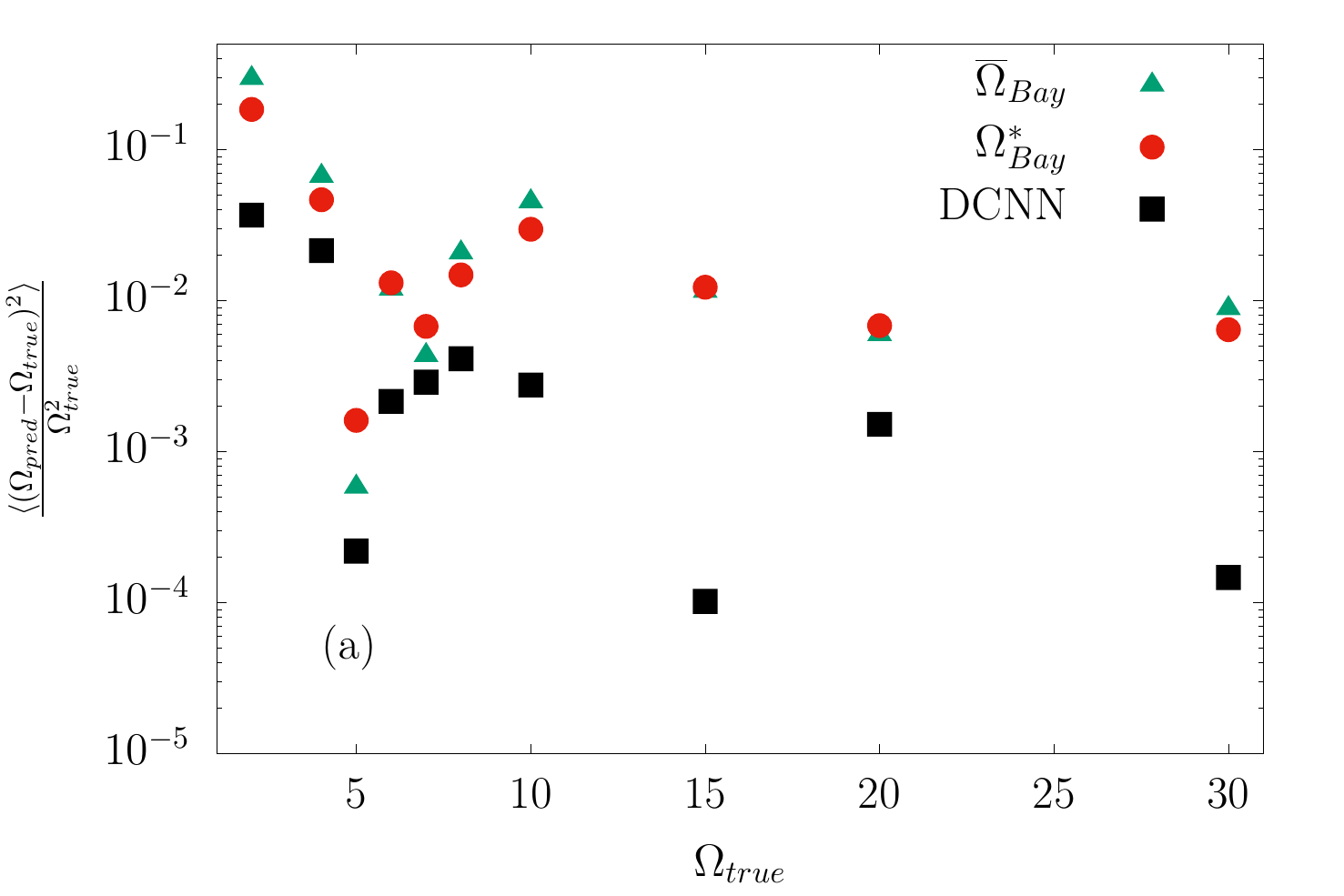}
\includegraphics[scale=0.55]{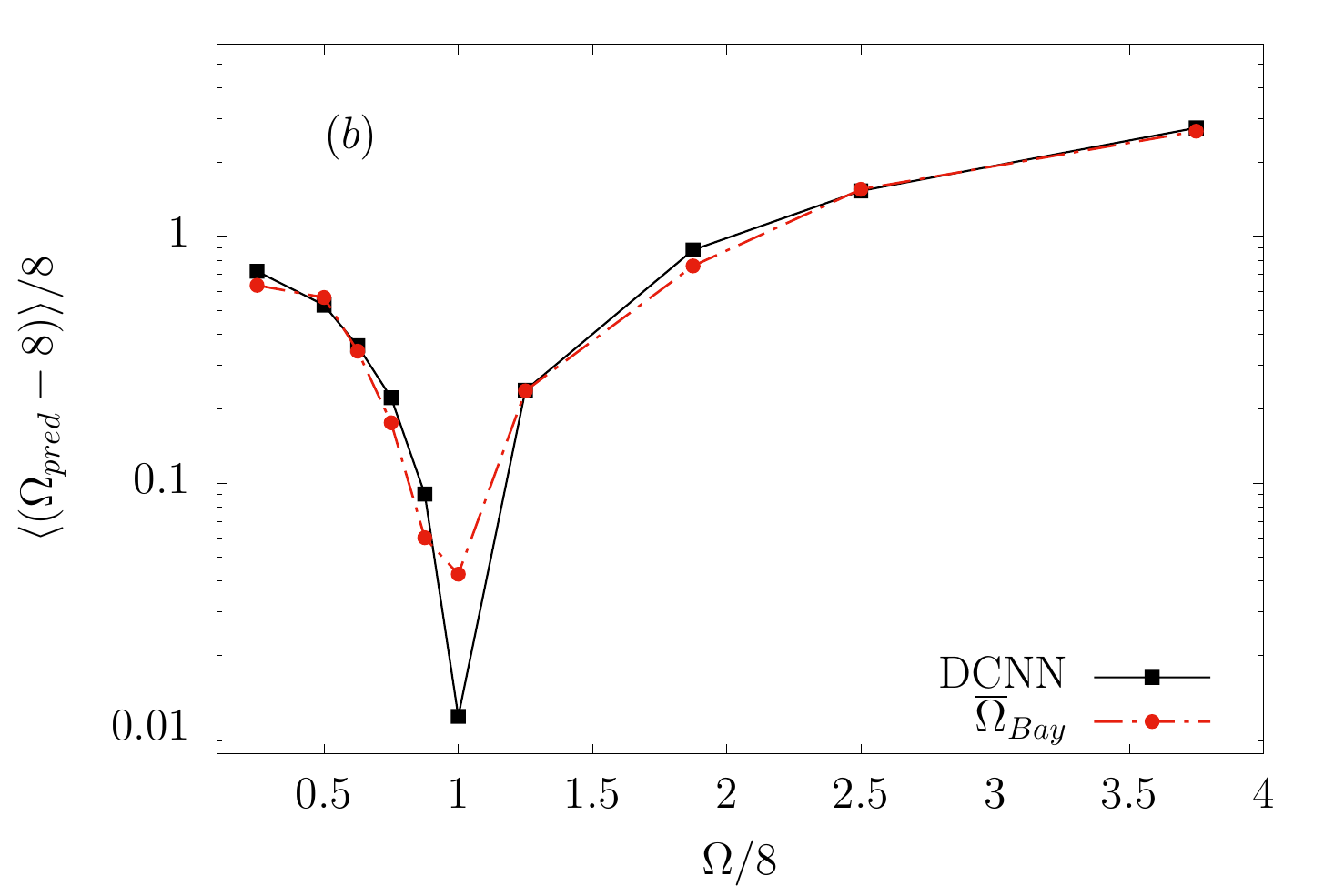}
\caption{\MB{(Panel a) Normalized mean squared error (MSE) between the predicted and true $\Omega$ values for each of the 10 classes considered. The average is taken over all validation planes of each simulations, while the predictions are obtained with DCNN (black squares), BI $\overline{\Omega}_{Bay}$ (green triangles) and BI $\Omega^*_{Bay}$ (red circles). All BI estimations are obtained using joint-likelihood $P( \langle u^2 \rangle ,\langle (\partial_x u)^2 \rangle | \Omega_i)$. (Panel b) Mean normalized distance between the mean predictions averaged over all planes in each simulation and the selected value of $\Omega=8$, plotted as a function of the simulation $\Omega$. Black squares solid line reports the DCNN predictions, red circles dashed line reports the BI predictions.}}
\label{fig:figure9}       % Give a unique label
\end{figure*}
The detailed comparison between DCNN and the best BI with joint likelihood on velocity module and gradients is presented in Fig.~\ref{fig:figure8}, where 10 different panels report the PDF of the DCNN and BI predictions, for all planes and all $\Omega$ of the validation set. Each PDF is normalized to its reference rotation value $\Omega_{true}$ such as $\Omega_{pred}/\Omega_{true} = 1$ indicates the perfect prediction. 
Predictions in the small rotation regime ($\Omega = 2,  4$) are the more difficult because no large-scale coherent structures are expected to appear in the flow and the resulting field is very little affected by the presence of rotation, so the predictions are easily confused among these two values. Comparing the two PDFs panels (a) and (b) we notice that while Bayes peaks always at the center between the two values, the DCNN is correctly peaked at the right value. \MB{In general for all the analyzed $\Omega$ (panels a-l) the DCNN accuracy is always very high and the larger errors happen only for a few cases on the $10k$ planes of the validation set.
The last comparison between DCNN and BI is presented in Fig.~\ref{fig:figure9}. Here in panel (a) there is a summary with the statistics of the different DCNN and BI predictions. To do so we measure the Mean Squared Error (MSE) prediction normalized to the target value,
\be
MSE = \frac{\langle (\Omega_{pred}-\Omega_{true})^2 \rangle}{\Omega_{true}^2}.
\label{eq:MSE}
\ee
From this figure we can see that DCNN is superior to the BI  in a consistent way for all rotation rates. On the other hand, in panel (b), we focus on the sensitivity of the two inferring tools to distinguish parameters used in different simulations. In particular we aim to identify from the analysis of all planes of each simulation the ones that belong to a specific rotation class, that in this case was chosen to be $\Omega=8$.
In Fig.~\ref{fig:figure9}(b) we measure the mean normalized distance,
\be
\frac{\langle \Omega_{pred}-8 \rangle}{8}.
\ee
between the prediction $\Omega_{pred}$ and the value of interest, $\Omega=8$, averaging over all planes from taken from the same simulation as a function of the different $\Omega$ values used in the simulations. We can see that the distance is sharply minimized when the planes come from the DNS with the chosen rotation $\Omega=8$. In other words, Fig.~\ref{fig:figure9}(b) highlights the sensitivity of the data-driven tools in classifying complex turbulent conditions from `ensembles' of different environments.}

\section{Occlusion}
\label{sect:Occlusion}

\begin{figure*}[h!]
\centering
\includegraphics[scale=0.6]{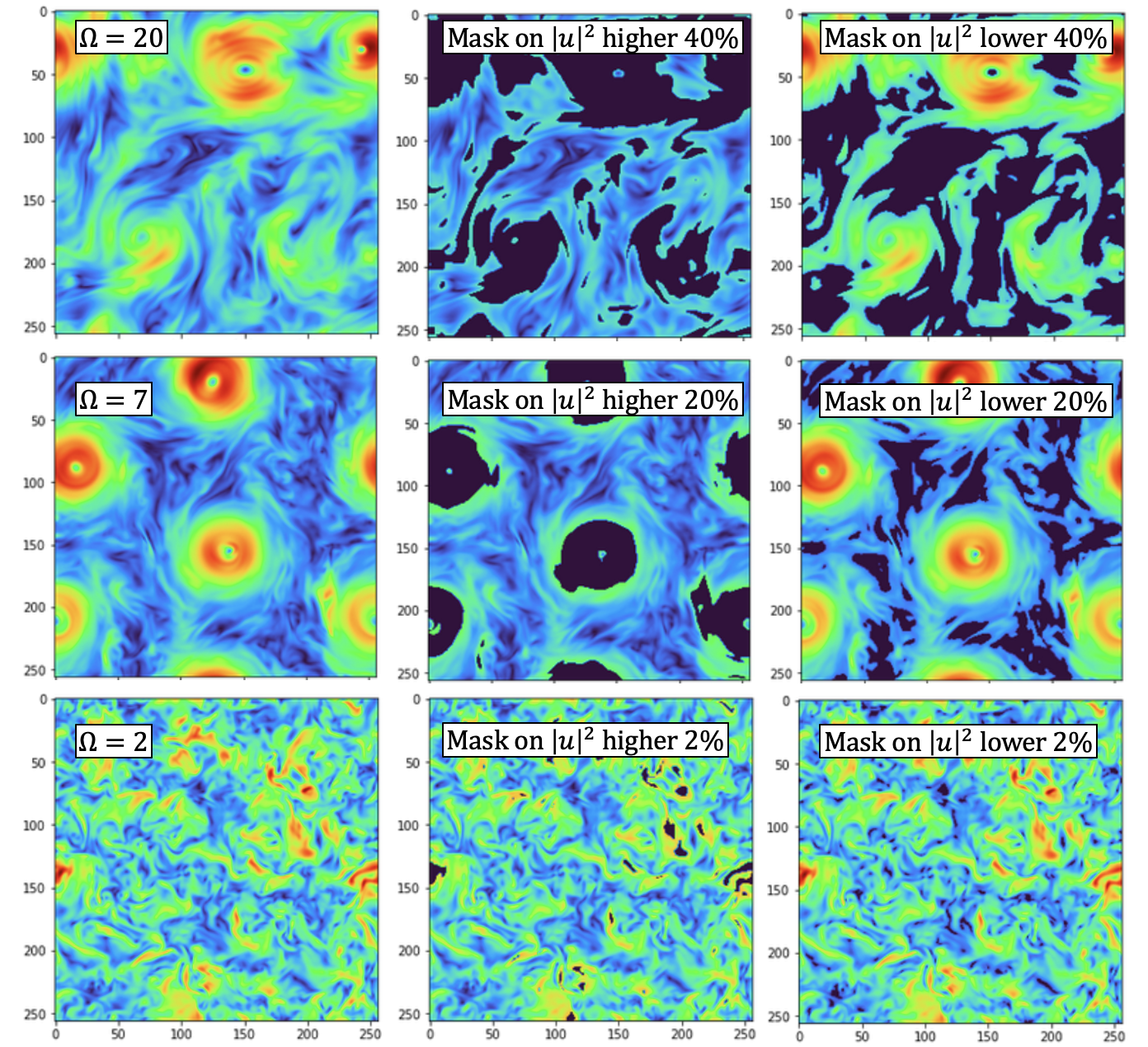}
\caption{Examples of real space occlusions defined on the high and low energy regions respectively for three different values of $\Omega$ and for three different percentage of points occluded. The three rows from top to bottom present visualizations for, $\Omega=20, 7$ and $2$, with a percentage of ablation of, $40\%, 20\%$ and $2\%$, on both the high and the low energy regions.}
\label{fig:figure10}       % Give a unique label
\end{figure*}

Here we present the ablation study on the input data aimed to perform a ranking on the importance of the flow features used by the DCNN to get a correct prediction, see~\cite{lellep2022interpreted} where a similar approach has been used on plane Couette flow to characterize the relevance of different regions for the prediction of relaminarisation events. The idea consists on keeping fixed the best DCNN trained model and to provide in input data where a fraction of the information were occluded such as to evaluate the deterioration of the inferring performance as a function of the different features occluded. As illustrated in Fig.~\ref{fig:figure10} the first type of ablation performed is done in real space and consists in applying a filter (zero-mask) over all high or low energy regions up to a given percentage of points in the plane. In this way we aim to occlude a particular spatial input region, characterized by its energy, in order to evaluate its relative importance for the network prediction. 
The real space occlusion fields can be defined as;
\begin{align}
u^{<}_o(\bx) &= \begin{cases}
u(\bx) &,\text{ if } |\bu|^2 < e_{h}\\
0 &, \text{ otherwise}
\end{cases}; \\
u^{>}_o(\bx) &= \begin{cases}
u(\bx)&, \text{ if } |\bu|^2 > e_{l}\\
0 &, \text{ otherwise}
\end{cases},
\end{align}
where $u_o^<$ is occluded field on the high energy regions, i.e. the mask removes all points where the energy is larger than $e_h$. The same idea holds for $u_o^>$, but now this field is masked on the low energy regions, namely on all points with energy lower than $e_l$. The two thresholds $(e_h, e_l)$, are chosen such that the percentage of points occluded is equal in the two cases.
In Fig.~\ref{fig:figure10} there is a visualization for both the high energy and low energy ablation for three different percentages of occlusion and three different rotation values.
\begin{figure*}[h!]
\centering
\includegraphics[scale=0.43]{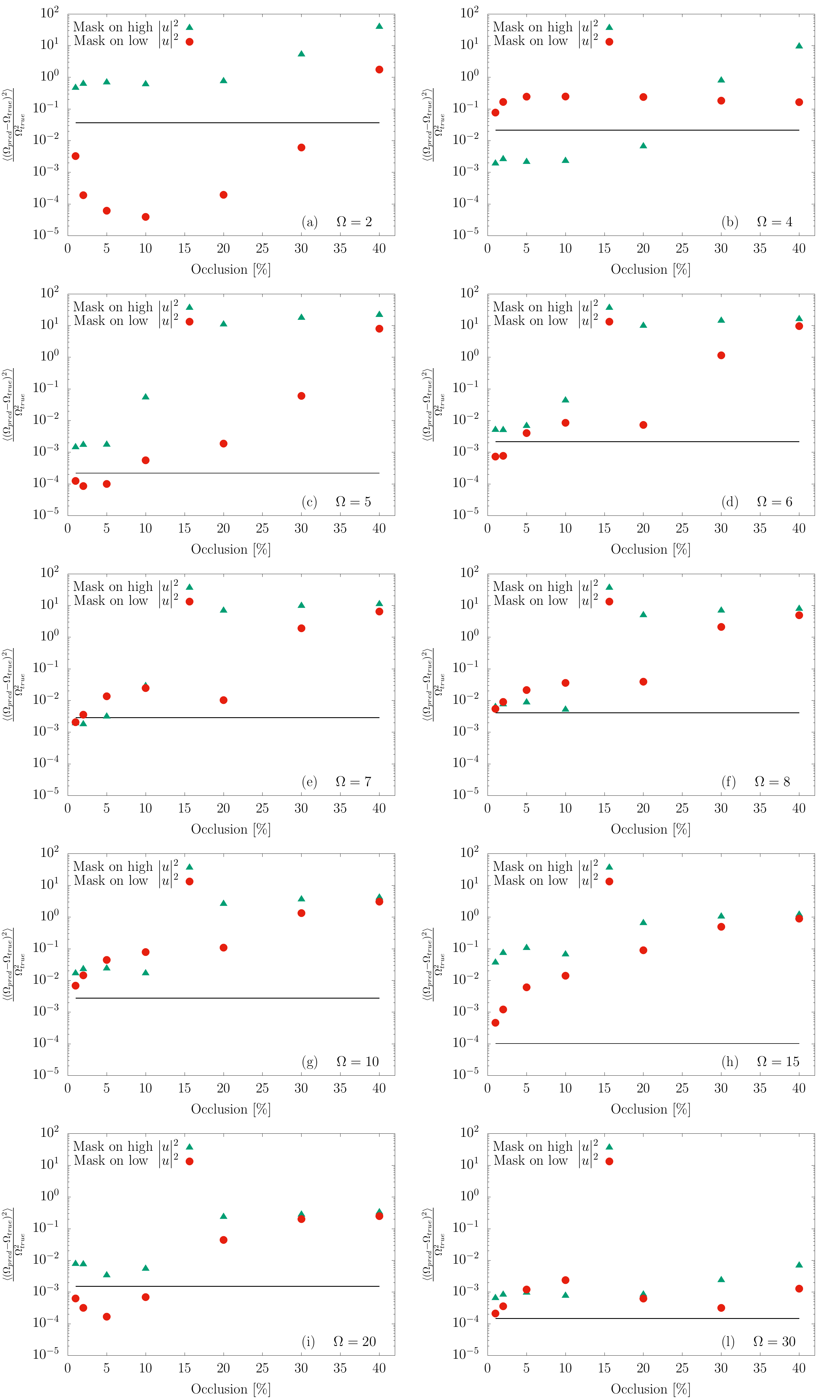}
\caption{MSE, as defined in Eq.~\eqref{eq:MSE}, obtained for the predictions on the occluded fields, varying the occlusion percentage for both masks applied to the higher energy (green triangles) and to the lower energy regions (red circles). Panels (a) and (b) present results for the weak rotation rates ($\Omega=2$ and $4$), while the remaining eight panels form (c) to (l) report results for the strong rotation rates where the fields are dominated by the intense vortical structures.}
\label{fig:figure11}       % Give a unique label
\end{figure*}

\begin{figure*}[h]
\centering
\includegraphics[scale=0.6]{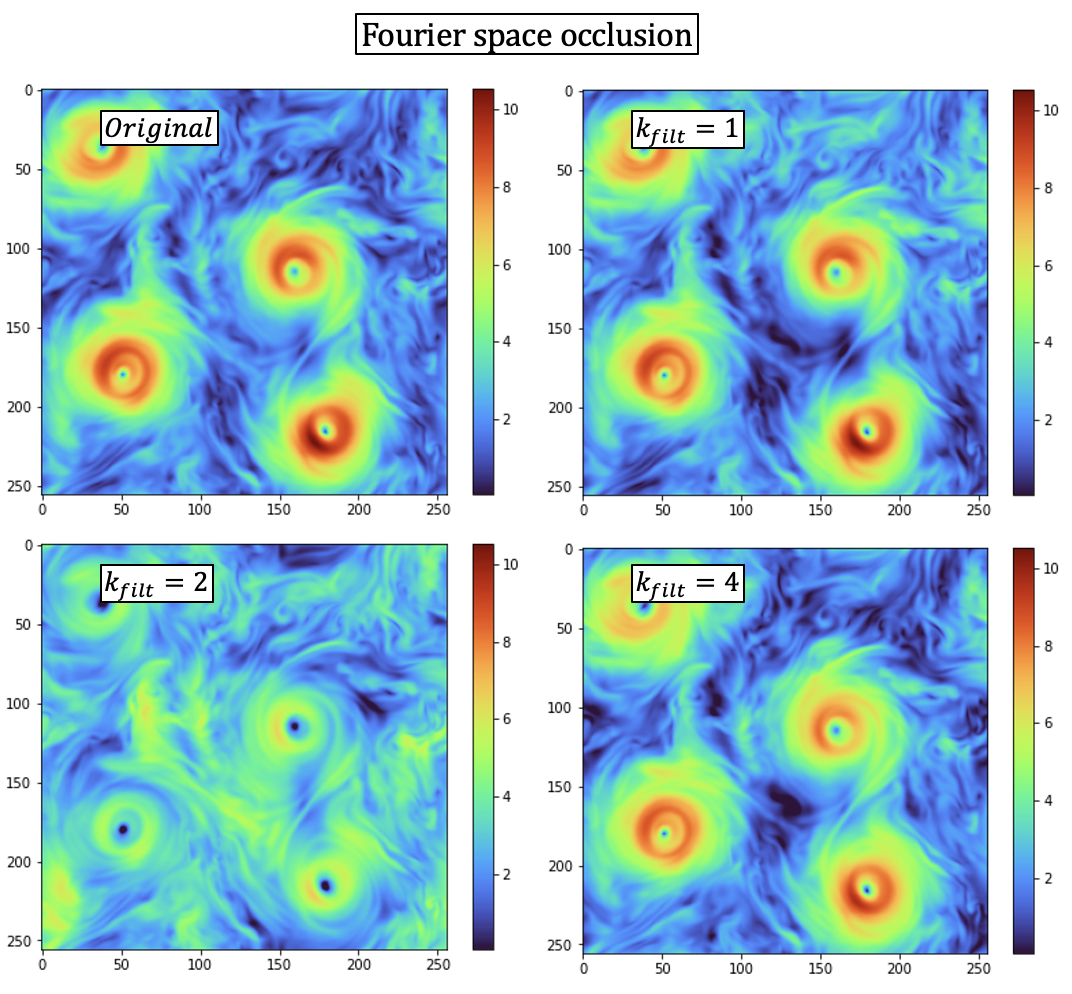}
\caption{Illustration of the effects of the Fourier type of occlusion applied at three different wavenumbers namely $k_{filt}=1,\,2, 4$, compared with the original flow configuration presented in the first top visualization.}
\label{fig:figure12}
\end{figure*}

\begin{figure*}[h]
\centering
\includegraphics[scale=0.55]{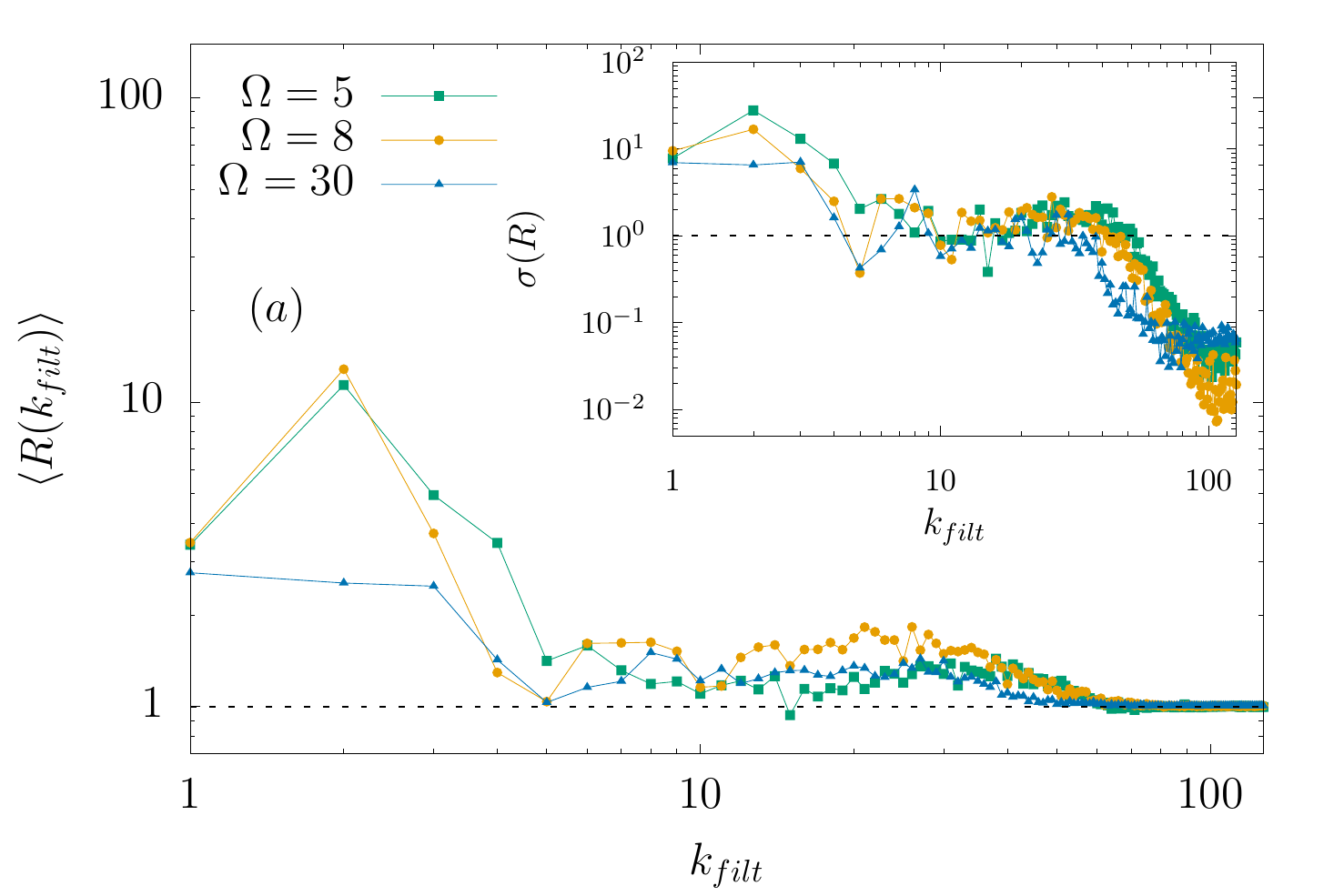}
\includegraphics[scale=0.55]{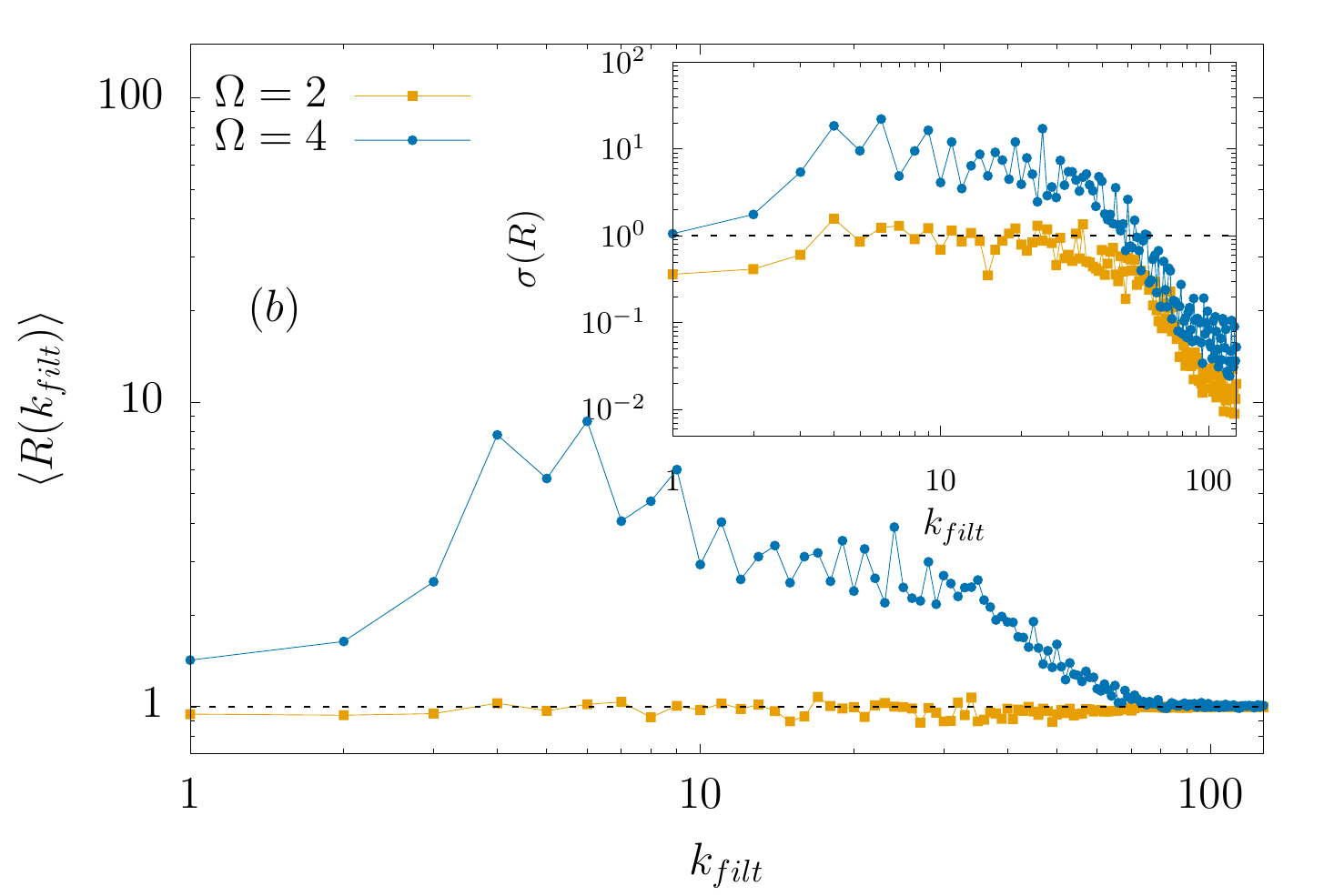}
\caption{Relevance $R(k_{filt})$, see Eq.~\eqref{eq:Rel}, resulting from the Fourier space ablation as a function of the filter wavenumber $k_{filt}$. Panel (a) shows results obtained at $\Omega=5,8$ and $30$, all in the inverse cascade regime. Panel (b) focuses on the two weak rotation rates $\Omega=2,4$. In both panels the main plot reports the mean relevance value ($\langle R\rangle$) while the inset reports its variance $\sigma(R)$. The horizontal dashed lines indicate the value $R(k_{filt})=1$, hence the value where the error on the occluded input is equal to the error on the full field.}
\label{fig:figure13}       % Give a unique label
\end{figure*}

To evaluate the importance of the occluded regions in terms of the network prediction, in Fig.~\ref{fig:figure11} we measure the MSE, as defined in Eq.~\eqref{eq:MSE}, obtained for the predictions on the occluded fields, varying the percentage of occlusion and comparing the ablation of high or low energy regions. Also in this case it is useful to distinguish between the weak and the strong rotation rates. The panels `a' and `b' show that for $\Omega=2$ the high energy regions are more important, and their ablation deteriorates the prediction. On the other hand for $\Omega=4$ the network focuses on the low energy regions which are found to be more important by this ablation study. All the remaining panels from `c' to `l' correspond to the strong rotation rates, and support in a consistent way that in the inverse cascade regime the intense vortical regions are more important for the correct classification. It is interesting to observe that for few cases, $\Omega=5, 6$ and $20$, the occlusion of a small percentage of the less energetic regions improved the DCNN prediction compared to the one obtained on full input.
The second type of occlusion, illustrated in Fig.~\ref{fig:figure12}, is a Fourier-space ablation. Here we remove all the degrees-of-freedom (DOF) contained in a specific circumference of radius $|\bk|=k_{filt}$ from the original plane in Fourier-space. Let us stress that the Fourier-space ablation is non-local in real space. Hence, all points of the occluded inputs are modified by the Fourier-space ablation.
It is also important to notice that when increasing the filter wavenumber, $k_{filt}$, the percentage DOF removed from the system becomes larger.
To quantify the relevance of a particular Fourier frequency $k_{filt}$ we have evaluated the relative error between the DCNN predictions evaluated on the occluded and on the full input, defined as;
\be
R(k_{filt}) = \frac{(\text{DCNN}[u_o(\bx, k_{filt})] - \Omega_{true})^2}{(\text{DCNN}[u(\bx)] - \Omega_{true})^2},
\label{eq:Rel}
\ee
where we have indicated as DCNN$[u(\bx)]$ the network prediction and as $u_o(\bx, k_{filt})$ the occluded field depending on the filter wavenumber. $R(k_{filt})>1$ indicates high-relevance for the occluded features because their removal deteriorates the prediction. On the contrary $R(k_{filt})<1$ suggest low relevance for the DOF at $k_{filt}$ whose removal improved the DCNN prediction.
In Fig.~\ref{fig:figure13} we report the results for the relevance $R(k_{filt})$ averaged over all planes of the validation set. Panel (a) shows results for a representative subset of the high rotation rates, $\Omega=5, 8$ and $30$, while panel (b) shows results for the two lower rotation rates. The main plot in both panels shows the mean relevance $\langle R(k_{filt}) \rangle$, as a function of the filter wavenumbers, averaged over all planes in the validation set of each $\Omega$.
As already observed from the real space occlusion, the role of the large scales depends on the rotation rate. When the rotation is strong, $\Omega \ge 5$, the relevance of the large and more energetic scales is preeminent. On the other hand, as expected, they are not relevant for the weak rotation rates.
Furthermore, with the Fourier space occlusion we have access to the role of the smaller scales and we found that the DCNN is sensitive to the occlusion on a wide range of scales up to $k_{filt} \sim 40$. This supports the idea that the neural network has learned to evaluate the complex multi-scale statistical properties to improve its prediction. Only in the particular case of $\Omega=2$ we have not observed any particular relevance at any scale.
In the inset of both panels (a) and (b) of Fig.~\ref{fig:figure13} we report the $R(k_{filt})$ variance as a function of the filter scale. Again the variance is evaluated over all planes of the validation set of each $\Omega$. From the variance we can highlight the presence of an exponential cutoff in the relevance of all wavenumbers larger than $k_{filt} \sim 40$. This observation tells us that such scales are dominated by viscous dissipation which is independent of the effect of rotation. 
In conclusion we have shown an attempt to use ablation to perform a features ranking on the flow properties more important for prediction of the DCNN. In our case we have found that the large scales swirling structures generated by rotation are generally more relevant for the correct classification of the high rotation rates. On the other hand we have seen that there is a full range of scales up to wavenumbers around $k_{filt}\sim 40$ that must be considered to improve the inferring prediction.

\section{Conclusions}
\label{sect:Conclusions}

We have studied the problem of inferring physical parameters of turbulent flows with purely data-driven tools without assuming any knowledge on the dynamical laws governing the systems. The case of study is 3d turbulence on a rotating frame with the goal of inferring the rotation frequency from the analysis of the flow velocity module on a 2d plane observed at a fixed time from the domain top. This regression problem, without considering the temporal dynamics of the flow, has not a clear solution even considering equation-driven approaches, and it represents an open question for theoretical physicists working on rotating flows. In this work we take advantage of the new ML tools and of a big pre-labeled dataset from high quality DNS~\cite{TurbRot} to tackle such problem. In particular we have trained a Deep Convolutional Neural Network in a supervised way to obtain a ML model able to solve this complex task. To compare the `supremacy' of DCNN results and to guide the reader towards a better understanding of the DCNN solution we have repeated the same game using standard Bayesian Inference (BI). For the latter estimation we have used two physically meaningful observables, moments of the velocity and of its gradients. In this way we have shown how the combination of more independent observables can allow to reach a better prediction, however simply increasing the number of observables would lead to a failure because of the difficulties to have a enough statistics to probe on the larger and larger dimensional space of the conditional probability. Performing an ablation study on the input data and fixing the best trained network we have shown that the DCNN is sensitive to the removal of small fraction of the input data, and it can be used to rank different flow features in terms of their importance for the prediction. Performing an ablation in real space we have shown that intense energy regions are relevant for a correct classification of high rotation rates. At the same time with a Fourier space ablation we have shown that DCNN is sensitive to changes in the input applied over a large range of wavenumbers, indicating the capability of the network to identify not only regions in space more relevant than other but also that it is measuring non-trivial and multiscale observables to improve its prediction.
As briefly discussed in the introduction, an unsupervised classification of the same dataset has been attempted using a GMM classification algorithm, however no relation between the classes identified and the rotation value has been found. Indeed GMM, as more in general k-means algorithms~\cite{lee1999learning}, are mainly sensitive to the number of vortexes inside each plane or to their spatial location and not to observables that are related to the rotation rate. For this reason a detailed analysis using this tool has been omitted in the paper. 
The classification of complex turbulent environments can be extended in several ways and has many real world applications. One example, consists in inferring of the intensification of extreme cyclones in the atmosphere~\cite{bhatia2019recent,xu2021deep}, where learning how to combine standard information with more complex data as satellites images can lead to better predictions of such phenomena. Another potential application is in self-navigation problem, where autonomous vehicles need to know the on-time flow conditions before to choose the best policy to optimize their navigation~\cite{biferale2019zermelo,orzan2022optimizing,novati2019controlled,reddy2016learning,reddy2018glider}. Also in the realm of turbulence modelling it would be beneficial to perform an ad-hoc fine-tuning of the closure model depending on the specific local turbulence condition to improve the training performances~\cite{maulik2019subgrid,pathak2020using,kochkov2021machine}. Our work show that DCNN can solve this kind of inferring problems with unprecedented accuracy still under the limitation of having a large quantity of good quality data.

\section*{Acknowledgement}

The authors thank Prof. Luca Biferale for inspiration and many useful discussions. This work was supported by the European Research Council (ERC) under the European Union’s Horizon 2020 research and innovation programme (Grant Agreement No. 882340), and by the project Beyond Borders (CUP): E84I19002270005 funded by the University of Rome Tor Vergata.

\section*{Data Availability Statements}
The datasets generated during and/or analysed during the current study are available in the Smart-Turb repository, \url{http://smart-turb.roma2.infn.it}.

\section*{Author contribution statement}
All authors contributed equally to the paper.

%ADD REFERENCES HERE
\bibliographystyle{unsrt}
\bibliography{biblio}

\begin{thebibliography}{10}

\bibitem{corbetta2021deep}
Alessandro Corbetta, Vlado Menkovski, Roberto Benzi, and Federico Toschi.
\newblock Deep learning velocity signals allow quantifying turbulence
  intensity.
\newblock {\em Science Advances}, 7(12):eaba7281, 2021.

\bibitem{alexakis2018cascades}
Alexandros Alexakis and Luca Biferale.
\newblock Cascades and transitions in turbulent flows.
\newblock {\em Physics Reports}, 767:1--101, 2018.

\bibitem{frisch1995turbulence}
Uriel Frisch.
\newblock {\em Turbulence: the legacy of AN Kolmogorov}.
\newblock Cambridge University Press, 1995.

\bibitem{Pope00}
Stephen~B. Pope.
\newblock {\em {Turbulent Flows}}.
\newblock Cambridge University Press, 2000.

\bibitem{davidson2011voyage}
Peter~A Davidson, Yukio Kaneda, Keith Moffatt, and Katepalli~R Sreenivasan.
\newblock {\em A voyage through turbulence}.
\newblock Cambridge University Press, 2011.

\bibitem{duraisamy2019turbulence}
Karthik Duraisamy, Gianluca Iaccarino, and Heng Xiao.
\newblock Turbulence modeling in the age of data.
\newblock {\em Annual Review of Fluid Mechanics}, 51:357--377, 2019.

\bibitem{biferale2019zermelo}
Luca Biferale, Fabio Bonaccorso, Michele Buzzicotti, Patricio Clark Di~Leoni,
  and Kristian Gustavsson.
\newblock Zermelo’s problem: Optimal point-to-point navigation in 2d
  turbulent flows using reinforcement learning.
\newblock {\em Chaos: An Interdisciplinary Journal of Nonlinear Science},
  29(10):103138, 2019.

\bibitem{novati2019controlled}
Guido Novati, Lakshminarayanan Mahadevan, and Petros Koumoutsakos.
\newblock Controlled gliding and perching through deep-reinforcement-learning.
\newblock {\em Physical Review Fluids}, 4(9):093902, 2019.

\bibitem{orzan2022optimizing}
N~Orzan, C~Leone, A~Mazzolini, J~Oyero, and A~Celani.
\newblock Optimizing airborne wind energy with reinforcement learning.
\newblock {\em arXiv preprint arXiv:2203.14271}, 2022.

\bibitem{garnier2021review}
Paul Garnier, Jonathan Viquerat, Jean Rabault, Aur{\'e}lien Larcher, Alexander
  Kuhnle, and Elie Hachem.
\newblock A review on deep reinforcement learning for fluid mechanics.
\newblock {\em Computers \& Fluids}, 225:104973, 2021.

\bibitem{colabrese2017flow}
Simona Colabrese, Kristian Gustavsson, Antonio Celani, and Luca Biferale.
\newblock Flow navigation by smart microswimmers via reinforcement learning.
\newblock {\em Physical review letters}, 118(15):158004, 2017.

\bibitem{reddy2016learning}
Gautam Reddy, Antonio Celani, Terrence~J Sejnowski, and Massimo Vergassola.
\newblock Learning to soar in turbulent environments.
\newblock {\em Proceedings of the National Academy of Sciences},
  113(33):E4877--E4884, 2016.

\bibitem{reddy2018glider}
Gautam Reddy, Jerome Wong-Ng, Antonio Celani, Terrence~J Sejnowski, and Massimo
  Vergassola.
\newblock Glider soaring via reinforcement learning in the field.
\newblock {\em Nature}, 562(7726):236--239, 2018.

\bibitem{scatamacchia2012extreme}
R~Scatamacchia, L~Biferale, and F~Toschi.
\newblock Extreme events in the dispersions of two neighboring particles under
  the influence of fluid turbulence.
\newblock {\em Physical review letters}, 109(14):144501, 2012.

\bibitem{buzzicotti2021inertial}
Michele Buzzicotti and Guillaume Tauzin.
\newblock Inertial range statistics of the entropic lattice boltzmann method in
  three-dimensional turbulence.
\newblock {\em Physical Review E}, 104(1):015302, 2021.

\bibitem{biferale2016coherent}
Luca Biferale, Fabio Bonaccorso, Irene~M Mazzitelli, Michel~AT van Hinsberg,
  Alessandra~S Lanotte, Stefano Musacchio, Prasad Perlekar, and Federico
  Toschi.
\newblock Coherent structures and extreme events in rotating multiphase
  turbulent flows.
\newblock {\em Physical Review X}, 6(4):041036, 2016.

\bibitem{buaria2020self}
Dhawal Buaria, Alain Pumir, and Eberhard Bodenschatz.
\newblock Self-attenuation of extreme events in navier--stokes turbulence.
\newblock {\em Nature communications}, 11(1):1--7, 2020.

\bibitem{yeung2015extreme}
PK~Yeung, XM~Zhai, and Katepalli~R Sreenivasan.
\newblock Extreme events in computational turbulence.
\newblock {\em Proceedings of the National Academy of Sciences},
  112(41):12633--12638, 2015.

\bibitem{maulik2019subgrid}
Romit Maulik, Omer San, Adil Rasheed, and Prakash Vedula.
\newblock Subgrid modelling for two-dimensional turbulence using neural
  networks.
\newblock {\em Journal of Fluid Mechanics}, 858:122--144, 2019.

\bibitem{pathak2020using}
Jaideep Pathak, Mustafa Mustafa, Karthik Kashinath, Emmanuel Motheau, Thorsten
  Kurth, and Marcus Day.
\newblock Using machine learning to augment coarse-grid computational fluid
  dynamics simulations.
\newblock {\em arXiv preprint arXiv:2010.00072}, 2020.

\bibitem{kochkov2021machine}
Dmitrii Kochkov, Jamie~A Smith, Ayya Alieva, Qing Wang, Michael~P Brenner, and
  Stephan Hoyer.
\newblock Machine learning--accelerated computational fluid dynamics.
\newblock {\em Proceedings of the National Academy of Sciences},
  118(21):e2101784118, 2021.

\bibitem{biferale2019self}
Luca Biferale, Fabio Bonaccorso, Michele Buzzicotti, and Kartik~P Iyer.
\newblock Self-similar subgrid-scale models for inertial range turbulence and
  accurate measurements of intermittency.
\newblock {\em Physical review letters}, 123(1):014503, 2019.

\bibitem{vallis2017atmospheric}
Geoffrey~K Vallis.
\newblock {\em Atmospheric and oceanic fluid dynamics}.
\newblock Cambridge University Press, 2017.

\bibitem{pedlosky1987geophysical}
Joseph Pedlosky et~al.
\newblock {\em Geophysical fluid dynamics}, volume 710.
\newblock Springer, 1987.

\bibitem{akyildiz2002wireless}
Ian~F Akyildiz, Weilian Su, Yogesh Sankarasubramaniam, and Erdal Cayirci.
\newblock Wireless sensor networks: a survey.
\newblock {\em Computer networks}, 38(4):393--422, 2002.

\bibitem{kalnay2003atmospheric}
Eugenia Kalnay.
\newblock {\em Atmospheric modeling, data assimilation and predictability}.
\newblock Cambridge university press, 2003.

\bibitem{buzzicotti_ocean_2021}
Michele Buzzicotti, Benjamin~A Storer, Stephen~M Griffies, and Hussein Aluie.
\newblock A coarse-grained decomposition of surface geostrophic kinetic energy
  in the global ocean.
\newblock {\em Earth and Space Science Open Archive}, page~58, 2021.

\bibitem{carrassi2008data}
Alberto Carrassi, Michael Ghil, Anna Trevisan, and Francesco Uboldi.
\newblock Data assimilation as a nonlinear dynamical systems problem: Stability
  and convergence of the prediction-assimilation system.
\newblock {\em Chaos: An Interdisciplinary Journal of Nonlinear Science},
  18(2):023112, 2008.

\bibitem{brunton2016discovering}
Steven~L Brunton, Joshua~L Proctor, and J~Nathan Kutz.
\newblock Discovering governing equations from data by sparse identification of
  nonlinear dynamical systems.
\newblock {\em Proceedings of the national academy of sciences},
  113(15):3932--3937, 2016.

\bibitem{bocquet2019data}
Marc Bocquet, Julien Brajard, Alberto Carrassi, and Laurent Bertino.
\newblock Data assimilation as a learning tool to infer ordinary differential
  equation representations of dynamical models.
\newblock {\em Nonlinear Processes in Geophysics}, 26(3):143--162, 2019.

\bibitem{smith1999transfer}
Leslie~M Smith and Fabian Waleffe.
\newblock Transfer of energy to two-dimensional large scales in forced,
  rotating three-dimensional turbulence.
\newblock {\em Physics of fluids}, 11(6):1608--1622, 1999.

\bibitem{mininni2009helicity}
Pablo~Daniel Mininni and A~Pouquet.
\newblock Helicity cascades in rotating turbulence.
\newblock {\em Physical Review E}, 79(2):026304, 2009.

\bibitem{biferale2021rotating}
Luca Biferale.
\newblock Rotating turbulence.
\newblock {\em Journal of Turbulence}, 22(4-5):232--241, 2021.

\bibitem{di2020phase}
P~Clark Di~Leoni, Alexandros Alexakis, L~Biferale, and M~Buzzicotti.
\newblock Phase transitions and flux-loop metastable states in rotating
  turbulence.
\newblock {\em Physical Review Fluids}, 5(10):104603, 2020.

\bibitem{zou1992optimal}
X~Zou, IM~Navon, and FX~LeDimet.
\newblock An optimal nudging data assimilation scheme using parameter
  estimation.
\newblock {\em Quarterly Journal of the Royal Meteorological Society},
  118(508):1163--1186, 1992.

\bibitem{ruiz2013estimating}
Juan~Jose Ruiz, Manuel Pulido, and Takemasa Miyoshi.
\newblock Estimating model parameters with ensemble-based data assimilation: A
  review.
\newblock {\em Journal of the Meteorological Society of Japan. Ser. II},
  91(2):79--99, 2013.

\bibitem{di2018inferring}
Patricio~Clark Di~Leoni, Andrea Mazzino, and Luca Biferale.
\newblock Inferring flow parameters and turbulent configuration with
  physics-informed data assimilation and spectral nudging.
\newblock {\em Physical Review Fluids}, 3(10):104604, 2018.

\bibitem{di2015spatio}
P~Clark di~Leoni, Pablo~J Cobelli, and Pablo~D Mininni.
\newblock The spatio-temporal spectrum of turbulent flows.
\newblock {\em The European Physical Journal E}, 38(12):1--10, 2015.

\bibitem{di2020synchronization}
Patricio~Clark Di~Leoni, Andrea Mazzino, and Luca Biferale.
\newblock Synchronization to big data: Nudging the navier-stokes equations for
  data assimilation of turbulent flows.
\newblock {\em Physical Review X}, 10(1):011023, 2020.

\bibitem{buzzicotti2020synchronizing}
Michele Buzzicotti and Patricio Clark Di~Leoni.
\newblock Synchronizing subgrid scale models of turbulence to data.
\newblock {\em Physics of Fluids}, 32(12):125116, 2020.

\bibitem{brenner2019perspective}
MP~Brenner, JD~Eldredge, and JB~Freund.
\newblock Perspective on machine learning for advancing fluid mechanics.
\newblock {\em Physical Review Fluids}, 4(10):100501, 2019.

\bibitem{DuraisamyARFM2019}
Karthik Duraisamy, Gianluca Iaccarino, and Heng Xiao.
\newblock Turbulence modeling in the age of data.
\newblock {\em Annual Review of Fluid Mechanics}, 51(1):357--377, 2019.

\bibitem{vinuesa2022enhancing}
Ricardo Vinuesa and Steven~L Brunton.
\newblock Enhancing computational fluid dynamics with machine learning.
\newblock {\em Nature Computational Science}, 2(6):358--366, 2022.

\bibitem{goodfellow2016deep}
Ian Goodfellow, Yoshua Bengio, and Aaron Courville.
\newblock {\em Deep learning}.
\newblock MIT press, 2016.

\bibitem{brunton2019data}
Steven~L Brunton and J~Nathan Kutz.
\newblock {\em Data-driven science and engineering: Machine learning, dynamical
  systems, and control}.
\newblock Cambridge University Press, 2019.

\bibitem{brunton2020machine}
Steven~L Brunton, Bernd~R Noack, and Petros Koumoutsakos.
\newblock Machine learning for fluid mechanics.
\newblock {\em Annual Review of Fluid Mechanics}, 52:477--508, 2020.

\bibitem{buzzicotti2021reconstruction}
Michele Buzzicotti, Fabio Bonaccorso, P~Clark Di~Leoni, and Luca Biferale.
\newblock Reconstruction of turbulent data with deep generative models for
  semantic inpainting from turb-rot database.
\newblock {\em Physical Review Fluids}, 6(5):050503, 2021.

\bibitem{brajard2019combining}
Julien Brajard, Alberto Carrassi, Marc Bocquet, and Laurent Bertino.
\newblock Combining data assimilation and machine learning to emulate a
  dynamical model from sparse and noisy observations: a case study with the
  lorenz 96 model.
\newblock {\em Geoscientific Model Development Discussions}, pages 1--21, 2019.

\bibitem{borra2021using}
Francesco Borra and Marco Baldovin.
\newblock Using machine-learning modeling to understand macroscopic dynamics in
  a system of coupled maps.
\newblock {\em Chaos: An Interdisciplinary Journal of Nonlinear Science},
  31(2):023102, 2021.

\bibitem{krizhevsky2012imagenet}
Alex Krizhevsky, Ilya Sutskever, and Geoffrey~E Hinton.
\newblock Imagenet classification with deep convolutional neural networks.
\newblock {\em Advances in neural information processing systems},
  25:1097--1105, 2012.

\bibitem{he2016deep}
Kaiming He, Xiangyu Zhang, Shaoqing Ren, and Jian Sun.
\newblock Deep residual learning for image recognition.
\newblock In {\em Proceedings of the IEEE conference on computer vision and
  pattern recognition}, pages 770--778, 2016.

\bibitem{ronneberger2015u}
Olaf Ronneberger, Philipp Fischer, and Thomas Brox.
\newblock U-net: Convolutional networks for biomedical image segmentation.
\newblock In {\em International Conference on Medical image computing and
  computer-assisted intervention}, pages 234--241. Springer, 2015.

\bibitem{redmon2016you}
Joseph Redmon, Santosh Divvala, Ross Girshick, and Ali Farhadi.
\newblock You only look once: Unified, real-time object detection.
\newblock In {\em Proceedings of the IEEE conference on computer vision and
  pattern recognition}, pages 779--788, 2016.

\bibitem{deusebio2014dimensional}
Enrico Deusebio, Guido Boffetta, Erik Lindborg, and Stefano Musacchio.
\newblock Dimensional transition in rotating turbulence.
\newblock {\em Physical Review E}, 90(2):023005, 2014.

\bibitem{marino2013inverse}
Raffaele Marino, Pablo~Daniel Mininni, Duane Rosenberg, and Annick Pouquet.
\newblock Inverse cascades in rotating stratified turbulence: fast growth of
  large scales.
\newblock {\em EPL (Europhysics Letters)}, 102(4):44006, 2013.

\bibitem{marino2014large}
Raffaele Marino, Pablo~Daniel Mininni, Duane~L Rosenberg, and Annick Pouquet.
\newblock Large-scale anisotropy in stably stratified rotating flows.
\newblock {\em Physical Review E}, 90(2):023018, 2014.

\bibitem{buzzicotti2018energy}
Michele Buzzicotti, Hussein Aluie, Luca Biferale, and Moritz Linkmann.
\newblock Energy transfer in turbulence under rotation.
\newblock {\em Physical Review Fluids}, 3(3):034802, 2018.

\bibitem{buzzicotti2018inverse}
Michele Buzzicotti, Patricio Clark Di~Leoni, and Luca Biferale.
\newblock On the inverse energy transfer in rotating turbulence.
\newblock {\em The European Physical Journal E}, 41(11):1--8, 2018.

\bibitem{seshasayanan2018condensates}
Kannabiran Seshasayanan and Alexandros Alexakis.
\newblock Condensates in rotating turbulent flows.
\newblock {\em Journal of Fluid Mechanics}, 841:434--462, 2018.

\bibitem{van2020critical}
Adrian van Kan and Alexandros Alexakis.
\newblock Critical transition in fast-rotating turbulence within highly
  elongated domains.
\newblock {\em Journal of Fluid Mechanics}, 899, 2020.

\bibitem{janai2020computer}
Joel Janai, Fatma G{\"u}ney, Aseem Behl, Andreas Geiger, et~al.
\newblock Computer vision for autonomous vehicles: Problems, datasets and state
  of the art.
\newblock {\em Foundations and Trends{\textregistered} in Computer Graphics and
  Vision}, 12(1--3):1--308, 2020.

\bibitem{maron2020learning}
Haggai Maron, Or~Litany, Gal Chechik, and Ethan Fetaya.
\newblock On learning sets of symmetric elements.
\newblock In {\em International Conference on Machine Learning}, pages
  6734--6744. PMLR, 2020.

\bibitem{pidhorskyi2018generative}
Stanislav Pidhorskyi, Ranya Almohsen, Donald~A Adjeroh, and Gianfranco Doretto.
\newblock Generative probabilistic novelty detection with adversarial
  autoencoders.
\newblock {\em arXiv preprint arXiv:1807.02588}, 2018.

\bibitem{tan2020efficientdet}
Mingxing Tan, Ruoming Pang, and Quoc~V Le.
\newblock Efficientdet: Scalable and efficient object detection.
\newblock In {\em Proceedings of the IEEE/CVF conference on computer vision and
  pattern recognition}, pages 10781--10790, 2020.

\bibitem{chouhan2020applications}
Siddharth~Singh Chouhan, Uday~Pratap Singh, and Sanjeev Jain.
\newblock Applications of computer vision in plant pathology: a survey.
\newblock {\em Archives of computational methods in engineering},
  27(2):611--632, 2020.

\bibitem{cheng2021fashion}
Wen-Huang Cheng, Sijie Song, Chieh-Yun Chen, Shintami~Chusnul Hidayati, and
  Jiaying Liu.
\newblock Fashion meets computer vision: A survey.
\newblock {\em ACM Computing Surveys (CSUR)}, 54(4):1--41, 2021.

\bibitem{feng2019computer}
Xin Feng, Youni Jiang, Xuejiao Yang, Ming Du, and Xin Li.
\newblock Computer vision algorithms and hardware implementations: A survey.
\newblock {\em Integration}, 69:309--320, 2019.

\bibitem{pathak2016context}
Deepak Pathak, Philipp Krahenbuhl, Jeff Donahue, Trevor Darrell, and Alexei~A
  Efros.
\newblock Context encoders: Feature learning by inpainting.
\newblock In {\em Proceedings of the IEEE conference on computer vision and
  pattern recognition}, pages 2536--2544, 2016.

\bibitem{rasmussen1999infinite}
Carl~Edward Rasmussen et~al.
\newblock The infinite gaussian mixture model.
\newblock In {\em NIPS}, volume~12, pages 554--560. Citeseer, 1999.

\bibitem{reynolds2009gaussian}
Douglas~A Reynolds.
\newblock Gaussian mixture models.
\newblock {\em Encyclopedia of biometrics}, 741:659--663, 2009.

\bibitem{HeGMM2011}
Xiaofei He, Deng Cai, Yuanlong Shao, Hujun Bao, and Jiawei Han.
\newblock Laplacian regularized gaussian mixture model for data clustering.
\newblock {\em IEEE Transactions on Knowledge and Data Engineering},
  23(9):1406--1418, 2011.

\bibitem{TurbRot}
Luca Biferale, Fabio Bonaccorso, Michele Buzzicotti, and Patricio Clark~di
  Leoni.
\newblock {TURB-R}ot. {A} large database of 3d and 2d snapshots from turbulent
  rotating flows. \url{http://smart-turb.roma2.infn.it}.
\newblock {\em arXiv:2006.07469}, 2020.

\bibitem{abadi2016tensorflow}
Mart{\'\i}n Abadi, Paul Barham, Jianmin Chen, Zhifeng Chen, Andy Davis, Jeffrey
  Dean, Matthieu Devin, Sanjay Ghemawat, Geoffrey Irving, Michael Isard, et~al.
\newblock Tensorflow: A system for large-scale machine learning.
\newblock In {\em 12th $\{$USENIX$\}$ symposium on operating systems design and
  implementation ($\{$OSDI$\}$ 16)}, pages 265--283, 2016.

\bibitem{srivastava2014dropout}
Nitish Srivastava, Geoffrey Hinton, Alex Krizhevsky, Ilya Sutskever, and Ruslan
  Salakhutdinov.
\newblock Dropout: a simple way to prevent neural networks from overfitting.
\newblock {\em The journal of machine learning research}, 15(1):1929--1958,
  2014.

\bibitem{dozat2016incorporating}
Timothy Dozat.
\newblock Incorporating nesterov momentum into adam.
\newblock 2016.

\bibitem{aster2018parameter}
Richard~C Aster, Brian Borchers, and Clifford~H Thurber.
\newblock {\em Parameter estimation and inverse problems}.
\newblock Elsevier, 2018.

\bibitem{efron2013250}
Bradley Efron.
\newblock A 250-year argument: Belief, behavior, and the bootstrap.
\newblock {\em Bulletin of the American Mathematical Society}, 50(1):129--146,
  2013.

\bibitem{mackay1992bayesian}
David~JC MacKay.
\newblock Bayesian interpolation.
\newblock {\em Neural computation}, 4(3):415--447, 1992.

\bibitem{lellep2022interpreted}
Martin Lellep, Jonathan Prexl, Bruno Eckhardt, and Moritz Linkmann.
\newblock Interpreted machine learning in fluid dynamics: explaining
  relaminarisation events in wall-bounded shear flows.
\newblock {\em Journal of Fluid Mechanics}, 942, 2022.

\bibitem{lee1999learning}
Daniel~D Lee and H~Sebastian Seung.
\newblock Learning the parts of objects by non-negative matrix factorization.
\newblock {\em Nature}, 401(6755):788--791, 1999.

\bibitem{bhatia2019recent}
Kieran~T Bhatia, Gabriel~A Vecchi, Thomas~R Knutson, Hiroyuki Murakami, James
  Kossin, Keith~W Dixon, and Carolyn~E Whitlock.
\newblock Recent increases in tropical cyclone intensification rates.
\newblock {\em Nature communications}, 10(1):1--9, 2019.

\bibitem{xu2021deep}
Wenwei Xu, Karthik Balaguru, Andrew August, Nicholas Lalo, Nathan Hodas, Mark
  DeMaria, and David Judi.
\newblock Deep learning experiments for tropical cyclone intensity forecasts.
\newblock {\em Weather and Forecasting}, 36(4):1453--1470, 2021.

\end{thebibliography}

\end{document}